\documentclass[12pt,preprint]{aastex}
\shorttitle{SN Ia Host Galaxies}
\shortauthors{************** et al.}

\begin{document}

\title{Chemistry and Star Formation in the Host Galaxies  \\
       of Type Ia Supernovae}

\author{Joseph S. Gallagher and Peter M. Garnavich}
\affil{Department of Physics, University of Notre Dame, 225 Nieuwland 
Science Hall, Notre Dame, IN 46556-5670}
\and
\author{Perry Berlind}
\affil{F. L. Whipple Observatory, 670 Mount Hopkins Road, P.O. Box 97, 
Amado, AZ 85645}
\and
\author{Peter Challis}
\affil{Harvard-Smithsonian Center for Astrophysics, 60 Garden Street, 
Cambridge, MA 02138}
\and
\author{Saurabh Jha}
\affil{Department of Astronomy, University of California, 601 Campbell 
Hall, Berkeley, CA 94720-3411}
\and
\author{Robert P. Kirshner}
\affil{Harvard-Smithsonian Center for Astrophysics, 60 Garden Street, 
Cambridge, MA 02138}

\begin{abstract}

We study the effect of environment on the properties of type Ia supernovae 
by analyzing the integrated spectra of 57 local type Ia supernova host 
galaxies.  We deduce from the spectra the 
metallicity, current star formation rate, and star formation history of the host 
and compare these to the supernova decline rates.  Additionally, we compare the host 
properties to the difference between the derived supernova distance and the 
distance determined from the best-fit Hubble law.  From this we investigate possible 
uncorrected systematic effects inherent in the calibration of type Ia supernova luminosities 
using light curve fitting techniques.  Our results indicate a statistically insignificant correlation 
in the direction higher metallicity spiral galaxies host fainter type Ia supernovae.  
However, we present qualitative evidence suggesting progenitor age 
is more likely to be the source of variability in supernova peak luminosities than 
is metallicity.  We do not find a correlation between the supernova decline rate 
and host galaxy absolute B magnitude, nor do we find evidence of a significant 
relationship between decline rate and current host galaxy star formation rate.  
A tenuous correlation is observed between the supernova Hubble 
residuals and host galaxy metallicities.  Further host galaxy observations will be needed to
refine the significance of this result.  Finally, we characterize the
environmental property distributions for type Ia supernova host galaxies 
through a comparison with two larger, more general galaxy distributions 
using Kolmogorov-Smirnov tests.  The results show the host galaxy metallicity 
distribution to be similar to the metallicity distributions of the galaxies of the 
NFGS and SDSS.  Significant differences are observed between the SN Ia 
distributions of absolute B magnitude and star formation histories and the 
corresponding distributions of galaxies in the NFGS and SDSS.  Among 
these is an abrupt upper limit observed in the distribution of star formation 
histories of the host galaxy sample suggesting a type Ia supernovae 
characteristic delay time lower limit of approximately 2.0 Gyrs.  Other 
distribution discrepancies are investigated and the effect on the 
supernova properties are discussed.

\end{abstract}

\section{Introduction}

Among the powerful tools that have come into prominence in the last decade 
in the field of cosmology, few have been as important in advancing the 
subject as type Ia supernovae (SNe Ia).  SNe Ia show variations in their peak 
luminosities and colors that correlate well with their light curve decay times 
\citep{P93,H96c,RPK96,Perl99} making SNe Ia the best distance indicators.  
Application of this empirical relationship to the low-z sample of galaxies enabled 
cosmologists to refine the measurement of the Hubble Constant with great precision 
\citep{Jha99,Fre01}. 

Realizing the potential for SNe Ia to act as accurate cosmological probes, researchers 
applied the technique to the high-z sample of galaxies \citep{Perl97,G98,Schmidt98}.  
This research has yielded evidence suggesting that our universe is in a state of 
accelerating expansion implying a form of dark energy whose nature we do not yet 
understand \citep{R98,Perl99}.  The goal of the ESSENCE project is to improve our 
understanding of this negative pressure by placing tight constraints on the cosmic 
equation of state through a study of $\sim$200 SNe Ia at intermediate redshifts.  The   
result will be a detailed map of the history of cosmic expansion with less than 2\% 
distance error in six redshift bins and the ability to constrain the equation of state to 
10\%.  Moreover, future studies will attempt to differentiate between a vacuum 
energy and other more exotic sources for the acceleration \citep{wang01,gajus05} 
that will push the limits of the SNe Ia reliability. Clearly, it is of paramount importance to 
understand any systematic uncertainties in the calibration of SNe Ia that could bias 
these cosmological measurements. 

Type Ia supernovae are identified by the presence of singly ionized silicon, 
magnesium, sulfur, calcium, and the conspicuous absence of hydrogen in 
their spectra.  Early statistical studies of type I supernovae and their host galaxies 
showed that the events, like core-collapse SNe, are associated with young stellar 
populations \citep{Oeml79,Cald81}.  However, unlike their core-collapse counterparts, 
type Ia supernovae are readily observed far from the spiral arms in spiral galaxies 
and in early-type galaxies with low star-formation rates.  These observations require 
a delay between formation and explosion that is long enough to allow for proper 
diffusion away from the spiral arms \citep{MC96} and imply a lower mass for the 
SNe Ia progenitors.  Specifically, SNe Ia are thought to be triggered 
by thermonuclear ignition in the core of a CO white dwarf near the Chandrasekhar 
mass limit (1.4 M$_{\sun}$).  Two models predict how the WD attains the mass 
necessary to initiate explosive burning.  The first model is the single degenerate 
model \citep{WI73,Nom82} that describes a binary system in which a WD accretes 
matter from a main sequence or red giant binary companion.  The second model, the 
double degenerate model \citep{Webb84,IT84}, describes the coalescing of two 
binary WDs whose combined mass exceeds the Chandrasekhar mass limit.  
Once the mass limit is reached, carbon ignites resulting in the outward 
propagation of a burning front from the WD core.

Detonation occurs if the burning front travels outward faster than the local sound 
speed, but such an explosion would convert most of the star to nickel and would 
leave too few intermediate mass elements compared to the observed spectra.  
A deflagration results when the flame front traverses the star subsonically, but this 
tends to produce too little kinetic energy to account for the observed velocities.   
A combination of these two scenarios, known as the delayed detonation model 
(DD model) \citep{Klv91,Yam92,WW94} appears to best fit the observations.  The 
DD model assumes flame front propagation velocity that begins as deflagration and 
subsequently transitions into detonation at a specific transition density.  Although 
the DD model has been able to match many of the features observed in SNe Ia, there 
remains many open questions. For example, what triggers the transition to a 
detonation and how does the WD build mass to reach the Chandrasekhar limit? 

These uncertainties reinforce the need to investigate systematic effects 
that can influence the luminosity-decline rate relation. One important effect is the 
possible evolutionary changes undergone by the stellar populations 
producing the supernova progenitors.  For example, systematic differences between 
the high-z host stellar populations and the local host stellar populations could contribute 
to an inherent difference between the peak luminosities of low-z events and those of the 
high-z events.  Fortunately, the local sample of galaxies provides such a wide range of 
host stellar environments that a study of these local environments can provide insight 
into environmental parameters that may correlate with redshift.

Theoretical models have shown that parameters such as progenitor mass and 
metallicity can have an effect on the luminosity and  light curve shape of the 
resultant supernova by influencing the relative CNO abundances in the white 
dwarf.  For the DD model, massive progenitors produce faint type Ia supernovae 
because of a low carbon fraction in the core \citep{HWT98,Um99}.  The carbon 
fraction is also lowered as the progenitor metallicity is increased resulting in less 
energetic explosions.   For the range of masses expected for CO white dwarfs,
lowering the carbon fraction is expected to affect the peak brightness of type Ia
events by about 20\%. The range in peak brightness due to progenitor metallicity 
variations is expected to be small unless the metal abundance is significantly 
higher than solar \citep{TBT03}.

In order for predictions such as these to be tested observationally, it would be 
necessary to analyze a large sample of Type Ia supernova host stellar populations 
covering a wide range of ages and metallicities against the parameters of the 
supernovae they produce.  Unfortunately, it is difficult to isolate and observe the 
specific stellar populations harboring the progenitor systems.  Moreover, a long 
delay between formation and explosion would blur the correlation between a 
SN characteristics and its present local environment.  Consequently, the majority 
of observational research in this topic has centered on the study of the integrated 
light from the SN Ia host galaxies.  An analysis of 
the integrated light has the added advantage of allowing for future comparisons 
with high-z host galaxies whose small angular size restrict the observations to 
integrated spectra.  

We characterize the SN environments through the spectroscopic study of 57 
type Ia supernova host galaxies.  We have two goals for this study.  The first is 
to take a direct look at the possible systematic effects that the host galaxy 
environment has on the SN Ia properties through an analysis of the 
interdependencies between host galaxy and SN Ia parameters.  Secondly, we 
take an indirect look at these systematics by comparing
our SN host sample with two larger, more general samples of galaxies - the 
galaxies of the Near Field Galaxy Survey and those of the Sloan Digital Sky 
Survey.  In \S 2 we introduce the host galaxy sample, detail the observing 
strategy, and present the data reduction process.  In \S 3, the spectroscopic 
results are presented, and the theoretical predictions are discussed in light 
of the results.  Finally, we summarize our conclusions.   

\section{Observations and Data Reduction}

\subsection{Observations}

The host galaxy spectra reported here were obtained with the FAST 
spectrograph \citep{Fab98} at the F.L. Whipple Observatory's 1.5 m 
Tillinghast telescope atop Mt. Hopkins in Arizona.  The data were taken 
during 13 nights between 1999 May through 2000 
September.  The seeing ranged from 1\arcsec\ to 2\arcsec\ throughout 
the survey.  The FAST spectrograph, with a 300 line mm$^{-1}$ reflection 
grating, allowed for 4,000 \AA\ coverage and a FWHM resolution of 
$\sim$6 \AA.  The slit was 3\arcsec\ wide and had an unvignetted length 
of 3\arcmin.  

The slit was aligned along each host galaxy's major axis to maximize the galactic light 
sampled.  The position angles for each major axis was determined using the Digital Sky 
Survey Plates (DSS).  The slit was offset to a distance matching the visible limit 
of the galaxy's minor axis on the DSS and the slit was scanned repeatedly 
across the galaxy during an exposure.  Exposure times ranged from 300 to 
1,200 s depending on the brightness of the target.  Seven of the target 
galaxies (NGC 2841, NGC 3368, NGC 3627, NGC 4526, NGC 4527, 
NGC 4536, and NGC 5005) had major axes that subtended angles larger than 
3\arcmin.  In these cases, we oriented the slit along the galaxy's minor axis and 
scanned along its major axis. It should be noted that light losses due to atmospheric 
refraction are expected to be minimal given our use of a relatively wide 3\arcsec\ slit and the 
fact that this slit was scanned across the entire visual extent of our galaxies, an extent 
that typically measured many times the width of our slit.

At the beginning and the end of each night's run, both 12 s flat exposures and 
bias exposures were taken.  Sky flats were taken to normalize the sensitivity along 
the slit.  Flux standard star exposures were obtained twice per night with the slit 
oriented along the star's parallactic angle \citep{Fil82}.  The standards were taken 
from the list given in \citet{Mass88}.  Preceding the observation of every object 
galaxy, we obtained a comparison spectrum of a He-Ne-Ar lamp for reference in 
the wavelength calibration. For every object galaxy, save a few that will be 
addressed later, three images were taken with identical exposure times.  Table 1 
details our galaxy sample and the relevant observational parameters pertaining to 
each.  Columns (1) and (2) give the common name of the target galaxy and the 
name of the supernova that it hosted, respectively. Column (3) shows the position 
angle of the slit for each object while the scan width is recorded in column (4).  
Column (5) gives the angular width of the extracted aperture for each host 
galaxy chosen to enclose all of the galactic light. 

\subsection{Data Reduction}

The data reduction performed during this study was conducted following the 
standard techniques within the IRAF\footnote{IRAF is the Image Reduction 
and Analysis Facility, a general purpose software system for the reduction 
and analysis of astronomical data. IRAF is written and supported by the IRAF 
programming group at the National Optical Astronomy Observatories (NOAO) 
in Tucson, Arizona. NOAO is operated by the  Association of Universities 
for Research in Astronomy (AURA), Inc. under cooperative agreement with the National Science 
Foundation} environment.  The data were both dark and bias subtracted.  
Each galaxy spectrum was flat-fielded to correct for pixel-to-pixel variability in 
the CCD detector.  Several pixels in each image were bad due to flaws on the CCD chip and 
had to be removed by interpolation.  This was accomplished using the FIXPIX routine 
in IRAF.  The acquisition of three identical spectra for each target galaxy allowed 
us to remove the majority of our cosmic rays by combining our images using the median 
parameter in the IRAF routine IMCOMBINE.  Any further anomalous pixels were removed 
individually using IMEDIT.  

There were two conditions
under which we were unable to remove cosmic rays in this fashion.  Firstly, in a few 
cases time constraints or poor atmospheric conditions prevented the acquisition of 3 
spectra for the given target galaxy.  In these situations, the cosmic rays were removed 
individually from those spectra we did obtain, and the images were averaged using 
IMCOMBINE.  Secondly, in order to determine whether those objects with three images 
could be combined successfully using the median parameter in IMCOMBINE, it was 
necessary to ensure both that the spatial axes of each spectrum were aligned, and that they 
each had comparable background levels.  Both of these tasks were accomplished using the 
IMPLOT routine to plot the average of several cuts along each image's respective 
spatial axis.  If the spatial axes were misaligned, they were shifted using the IMSHIFT 
routine in IRAF.  Occasionally, short term atmospheric variability resulted in evident 
variations observed in the continuum levels of the three image set.  If it was 
discernible which image(s) was bad, then that image(s) was removed and cosmic ray removal 
proceeded as detailed above on the remaining spectra.  On the other hand, if the anomalous 
image(s) was not evident, then the cosmic rays were removed individually 
from each image, aperture extraction was performed, and the extracted apertures were 
averaged using IMCOMBINE.

The next step was to extract a one-dimensional spectrum from each combined image using 
the APALL routine in IRAF.  The apertures were fitted interactively within IRAF and chosen to 
span a region on the spatial axis that extended slightly into the sky portion of the image on either 
side of the galaxy spectrum.  In this way we ensured the inclusion of nearly 100$\%$ of the galactic light.
However, attempts were made to avoid the inclusion of foreground stars in the aperture.  The sky levels 
and trace were defined interactively using APALL with a linear fit for the former and a third order cubic 
spline for the latter.  Wavelength and flux calibration proceeded using the standard techniques within IRAF.  

Following the flux calibration, a telluric absorption correction was performed on those galaxy spectra 
containing relevant emission lines (i.e. H$\alpha$ and SII) that have been redshifted into 
the B-band (6860-6890 \AA\ ) and beyond. Next, the spectra were dereddened to account for local reddening 
due to Galactic extinction.  This was done using the routine DEREDDEN in IRAF.  In each case 
a value of 3.0 was taken for the total to selective visual absorption ratio, \textit{R}.  Furthermore, 
the value of the color excess, \textit{E(B-V)}, was chosen for each galaxy direction to correspond 
to that which is stated by the NASA/IPAC Extragalactic Database (NED).  These color excess 
values were calculated from COBE, the IRAS maps, and the Leiden-Dwingeloo maps of HI 
emission.  Finally, the galaxy spectra were Doppler corrected using the routine DOPCOR with 
redshifts obtained from NED.

\subsection{Line Strengths}

Following reduction the spectral properties were analyzed through the identification and 
subsequent line profiling of various relevant spectral lines.  In each case the line strengths 
were recorded using the SPLOT routine within IRAF.  Gaussian line profiles were fit for 
each emission line individually with the primary source of error originating in the 
continuum definition.  If appropriate, a boxcar smoothing algorithm was applied interactively 
allowing for more accurate continuum definition.  We obtained both equivalent width (EW) 
and line fluxes for [OII] $\lambda$3727 (our resolution was insufficient to resolve the [OII] 
doublet), H$\beta$ $\lambda$4861, [OIII] $\lambda$4959, [OIII] $\lambda$5007, [OI] $\lambda$6300, 
[NII] $\lambda$6548, H$\alpha$ $\lambda$6562, [NII] $\lambda$6584, [SII] $\lambda$6717, [SII] 
$\lambda$6731.  The equivalent widths measured in Angstroms are shown in Table 2 while 
emission-line fluxes in units of 10$^{-14}$ ergs cm$^{-2}$ s$^{-1}$ are given in Table 3, 
respectively.  

\section{Results}

\subsection{Host Galaxy and Supernova Parameterization}

Here we describe the parameters which characterize the galaxies in the SN Ia host 
sample.  The galactic parameters are given in Table 4.  Columns (3)-(6) are observed 
while columns (7) and (8) are derived parameters.  Column (1) lists the galaxies in our 
sample while column (2) gives each galaxy's hosted supernova.  The absolute B 
magnitudes of each host along with their corresponding errors are recorded in columns 
(3) and (4), respectively.  The vast majority of magnitudes were calculated from distances 
derived from their respective redshifts.  In a few cases, the potential for uncertainty was 
heightened due to the low recessional velocity of the host galaxy.  Therefore, other distance 
measurements from the literature were employed for these cases when possible.  Cepheid 
based distance moduli were found for NGC 3368, NGC 3627, NGC 4639, and NGC 4536 
from the HST Key Project published in \citet{Gib00}.  The distance to NGC 4526 was 
determined by \citet{H96b} using the Surface brightness fluctuations/planetary nebula 
luminosity function and published in \citet{H00} (hereafter H00).  All magnitudes were 
corrected to correspond to a Hubble constant of 72 km s$^{-1}$ Mpc$^{-1}$.  
Column (5) lists the morphological types according to NED while column (6) shows the 
H$\alpha$ luminosity for each host galaxy.  

The derived galactic parameters were metallicity and birthrate parameter \textit{b} 
\citep{Sc86}.  These are shown in columns (7) and (8), respectively.  For those
host galaxies with distinguishable emission lines, we determined the metallicities 
from our emission line flux measurements using the models detailed in \citet{KD02}.  
The paper provides a series of line strength diagnostic diagrams with various 
dependences on both metallicity and the local ionization parameter, \textit{q}.  One 
first estimates an initial metallicity through a diagnostic that varies little with \textit{q}. 
The initial value is then used to pin down the value of the ionization parameter 
through a diagnostic with strong dependences on both metallicity and the ionization 
parameter.  Successive iterations ultimately provide the best estimate of the galaxy's 
metallicity.  For full details see \citet{KD02}.  Extinction correction was applied for those
galaxies with measurable Balmer emission using the Whitford reddening curve as 
paramaterized by \citet{MM72}.  We were unsuccessful in obtaining 
metallicity estimates for galaxies with weak emission.  Furthermore, our signal 
to noise was insufficient to provide accurate absorption line strengths needed for 
an absorption line metallicity estimate.  However, three galaxies from our sample 
had metallicities measured in H00 that we used in our analysis.

The final host galaxy parameter was the Scalo \textit{b} parameter \citep{Sc86}.  
The Scalo \textit{b} parameter is the ratio of the current star formation rate to the 
average star formation rate of the past.  The parameter was determined by 
interpolation of the plot given in Fig.3 from \citet{KTC94}(hereafter KTC94).  
The plot shows the dependence of \textit{b} on EW(H$\alpha$ + [NII]) as dictated 
by the exponential plus burst model detailed in KTC94.    

Our SNe were characterized according to the parameters shown in Table 5.  Once 
again, Column (1) and (2) give the galaxy ID and SN name, respectively.  Column (3) 
and column (4) give the decline rate parameter, $\Delta$m$_{15}$(B), and its 
corresponding error, respectively.  $\Delta$m$_{15}$(B) is defined as the change 
in apparent magnitude from maximum to 15 days after maximum light in the 
supernova rest frame.  It acts as a convenient, reddening free, indicator of luminosity.  
Finally, column (5) shows the Hubble residuals for the SN set.  Assuming that type 
Ia supernovae are perfect standard candles, the 
extinction and light curve corrected absolute magnitudes for the SNe should 
be identical.  The Hubble residual for an individual SN is then defined as the 
deviation of its light curve and color corrected absolute magnitude from the average light 
curve corrected magnitude in the SN set.  Our magnitudes originate from the set of 80 
Hubble flow SNe published in \citet{Jha02}.  

\subsection{SNe Ia and Host Galaxy Correlations}

It is the goal of this study to investigate the correlations between type Ia properties and 
their global host galaxy parameters.  Some of these correlations have been explored 
observationally in the past.  \citet{H96b} reported that the most luminous SNe Ia tend 
to be hosted by late type, spiral galaxies.  The same behavior is seen in Figure 1 
where we have replicated the morphological classification versus decline rate plot of 
\citet{H96b} for a large sample of host galaxies.  The data was compiled from the SNe 
described in \citet{Phil99}, \citet{Jha02}, \citet{Rie99}, and  \citet{kevin}.  
The vertical lines in Figure 1 represent the average decline rates for the SNe in late 
type and early type galaxies, respectively.  They confirm the results of Hamuy et al. that 
the slower declining, more luminous supernovae are hosted by late type galaxies.  
However, it is important to note when grouping host galaxies by their morphological type 
that such a grouping does not necessarily imply the members of a common class 
possess similar physical characteristics such as metallicity and star formation histories.  
For example, Figure 1 highlights NGC 2841 and NGC 0632 which are categorized as 
Sa and S0 galaxies, respectively.  However, NGC 2841 is a galaxy with none of the usual 
emission features typically observed in spiral galaxies.  Moreover, NGC 0632, although 
tentatively labeled an S0 galaxy by NED, shows strong emission indicative of a starburst 
galaxy.  This shows that the gross morphology provides a helpful, though incomplete 
picture of the host properties and that a more detailed host galaxy characterization is 
necessary.

\subsubsection{Metallicity}

Theoretical studies conducted by \citet{HWT98,Um99,HNUW00}(hereafter HNUW00) 
along with analytical analysis by \citet{TBT03}(hereafter TBT03) have suggested that the 
initial metallicity of the Ia progenitor can have a small, but possibly significant effect 
(particularly at high metallicity) on the luminosity of the resultant supernova.  HNUW00 
pointed out that the metallicity of the progenitor on the main sequence can affect the 
mass of the interior C/O core left behind as a WD, and ultimately affect the amount 
of $^{56}$Ni produced in the explosion and the peak luminosity of the SN Ia.  TBT03 
\begin{figure}[t]
\begin{center}
\figurenum{1}
\includegraphics[width=0.7\textwidth]{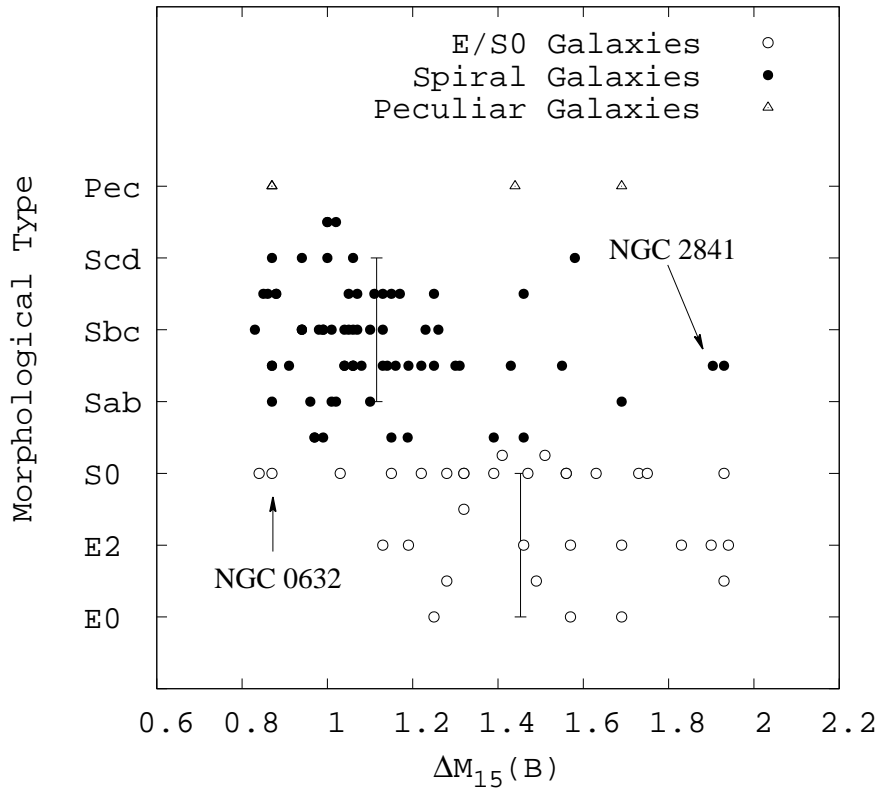}
\caption{Morphological Type vs. Decline Rate of SN.  Highlighted are NGC 2841, 
a categorized Sa galaxy with spectral features of an elliptical, and NGC 0632, a 
categorized S0 with the strong emission typically seen in late type galaxies. 
Vertical markers highlight the average decline rates for both early and late type 
host galaxies.}
\end{center}
\end{figure}   
analytically demonstrated that a factor of three variation in progenitor metallicity results 
in an $\sim$25$\%$ variation is the mass of $^{56}$Ni ejected during the Ia event.  If one 
allows for the sedimentation of $^{22}$Ne, then the variation can be as high as 50$\%$.  
Furthermore, \citet{Um99} suggested that the carbon mass fraction in a CO white dwarf 
is dependent on the metallicity of the environment in which it was formed.  They further 
proposed that the observed diversity in SNe Ia brightness is a consequence of this 
phenomenon with the smaller progenitor carbon fractions leading to dimmer supernova.  
Under the assumption that higher galactic metallicity is proportional to the 
average progenitor metallicity, we set out to investigate these theoretical 
results through observation. 

Figure 2 shows the relationship between host galaxy metallicity and SNe Ia decline rate 
for our sample with a distinction drawn between elliptical and spiral galaxies.  Two 
\begin{figure}[h!]
\begin{center}
\figurenum{2}
\includegraphics[width=0.7\textwidth]{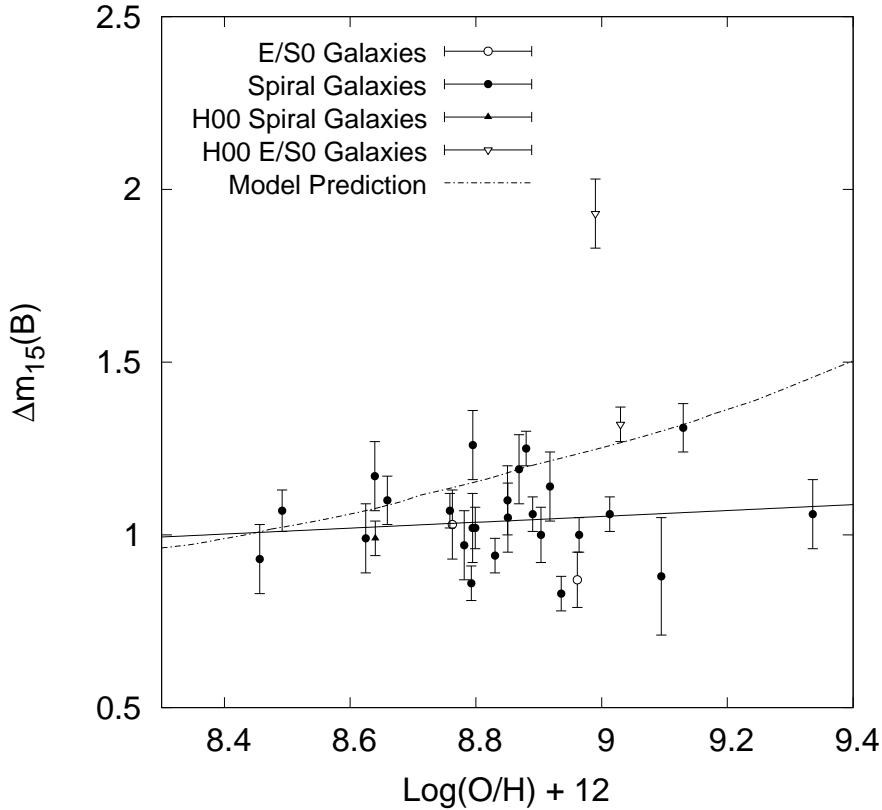}
\caption{Decline Rate dependence on Metallicity.  SN Ia sample of late 
types show a tendency for high metallicity galaxies to host fainter SNe 
Ia at the 70$\%$ confidence level.  Addition of H00 galaxies further 
suggests a decrease in SN brightness with increased metallicity from 
early to late type galaxies.   The solid line is a linear best fit to the data 
while the broken line represents the predicted trend governed by the 
studies of TBT03, \citet{Hof03}, and \citet{garn04}.}
\end{center}
\end{figure}
ellipticals and one spiral galaxy have been included with metallicities given in H00.  
A regression line is fitted to our host sample of spiral galaxies, and we find a 
small correlation suggesting that higher metallicity spiral host galaxies produce faster 
declining, less luminous SNe.  However, a Monte Carlo simulation places this correlation 
at only the 70$\%$ confidence level.  The simulation involved generating 25 evenly 
distributed metallicities between 8.3 and 9.4, assigning each one a $\Delta$m$_{15}$(B) 
from our data, and determining after a number of trials the probability of obtaining a best fit 
slope greater than or equal to the absolute value of the slope seen in Figure 2.  This does not 
suggest that metallicity has a great affect on the luminosity of type Ia supernovae.

A comparison between the late type galaxies of H00 and our spiral galaxies at the same metallicity 
shows a wide dispersion in decline rate at a fixed metallicity likewise suggesting a weak dependence of 
Ia decline rate on the environment metallicity.  However, the presence of two metal rich 
ellipticals from H00 that hosted SNe Ia with a higher decline rate 
on average than our spirals could hint at an overall increase in $\Delta$m$_{15}$(B) with 
metallicity across the full Hubble sequence.  Such a trend would support the predictions 
made by the analytical models of TBT03.  We use the DD numerical models of \citet{Hof03} and the 
empirical relations of \citet{garn04} to convert the predictions of TBT03 to the observed parameter.  
The predicted relation is shown in Figure 2 (broken line).  Oxygen abundances were converted into iron 
abundances using the [O/Fe] to [Fe/H] relation predicted by the three-component mixing models of 
\citet{QW01}.  $^{56}$Ni masses were calculated for the iron abundances according to the 
analytical model of TBT03.  The decline rate lower limit in the plotted range is set by the 
M$_{Ni}$ vs. metallicity relation presented in TBT03.  We assumes a fiducial SN Ia 
M$_{Ni}$ production of approximately 0.64M$_{\sun}$.  Varying this fiducial mass acts to vary the 
low metallicity decline rate limit.  Interpolation of the M$_{V}$ 
vs. M$_{Ni}$ plot in \citet{Hof03} yielded corresponding M$_{V}$ for each $^{56}$Ni mass.  Finally, 
we found corresponding decline rates through the empirical M$_{V}$ vs. $\Delta$m$_{15}$(B) relation 
presented in \citet{garn04}.  This predicted curve is not meant to provide for a detailed comparison with the observations, but rather to convey the general metallicity-decline rate trend predicted by TBT03.   The curve implies a minimal dependence of type Ia SN luminosity on metallicity for metallicities below solar.  However, as progenitor metallicity increases well above solar, the predicted 
dependence becomes steeper resulting in significantly fainter SNe Ia.

Figure 3 shows the projected galactocentric distances (PGD) of type Ia 
supernovae versus $\Delta$m$_{15}$(B).  The SNe were compiled from 
the list presented in \citet{Phil99}, \citet{Jha02}, \citet{Rie99}, and \citet{kevin}.  
Projected offsets were obtained from the Harvard CfA List of 
Supernovae\footnote{http://cfa-www.harvard.edu/cfa/ps/lists/Supernovae.html} and 
the Central Bureau for Astronomical Telegrams (CBAT).  Hubble flow luminosity 
distances were calculated from the SN redshift assuming cosmological parameters 
H$_{0}$ = 72 km s$^{-1}$ Mpc$^{-1}$, $\Omega$$_{M}$ = 0.28 and 
$\Omega$$_{\Lambda}$ = 0.72 while non-Hubble flow distances were estimated 
using the SNe luminosities.  A more even distribution in decline rates is observed for supernovae hosted 
by elliptical galaxies than those hosted by their smaller 
\begin{figure}[t]
\begin{center}
\figurenum{3}
\includegraphics[width=0.7\textwidth]{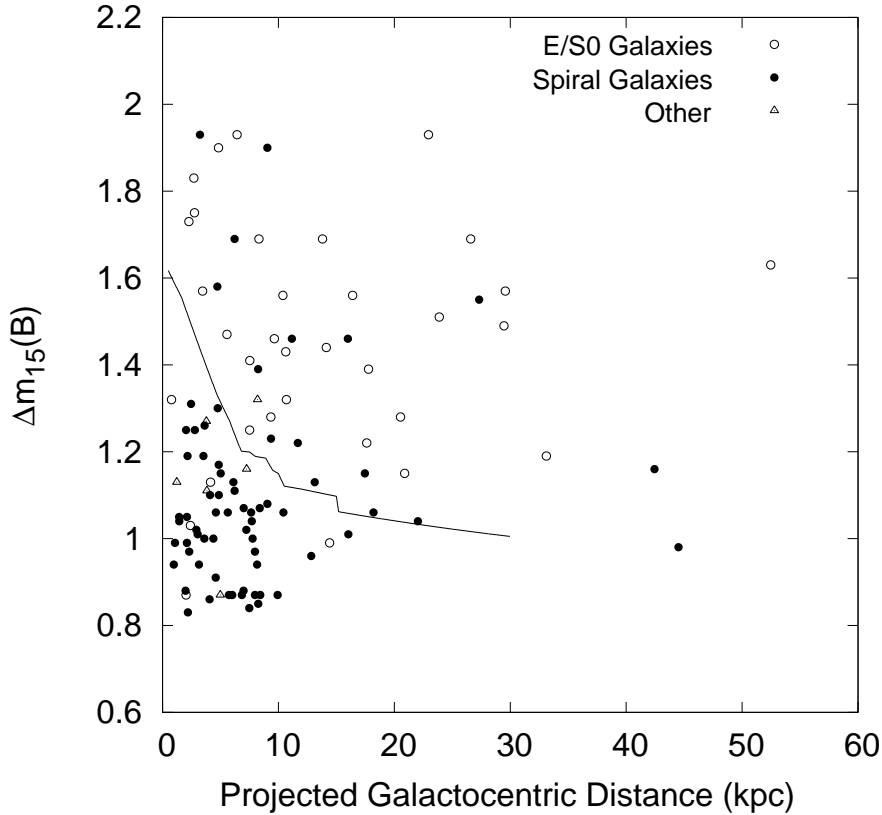}
\caption{Type Ia supernovae decline rate versus projected galactocentric distance.  Solid line represents
the predicted decline in $\Delta$m$_{15}$(B) with PGD due to a decrease in metallicity with increasing
PGD \citep{and04}.}
\end{center}
\end{figure}
spiral counterparts.  The relevance of this plot to metallicity becomes clear when we 
compare it to the expected metallicity gradient across a typical spiral galaxy.  Recent 
results by \citet{and04} have indicated that there is a drop of 0.6 dex in [Fe/H] across 
the disk of the Milky Way from approximately 4.0 kpc to 16 kpc.  Assuming the Milky 
Way to be adequately representative of a typical spiral galaxy, we can find a 
theoretical relation between the SN PGD and its decline rate using the methods detailed 
above.   The relation is potted in Figure 3, and it suggests that amongst SNe hosted 
by spiral galaxies, the fainter events are predicted to reside nearest the galactic center 
- a prediction that stands in contrast to the observations in Figure 3 showing the 
brighter events clustering at low PGDs. 

We also compared the SN hosts with two larger sets of galaxies in the hopes of 
shedding light on possible systematic selection effects or, more interestingly, 
evolutionary effects present in the discovery of SNe Ia.  The first set was the
Near Field Galaxy Survey sample (NFGS).  The NFGS is a collection of integrated 
and nuclear spectroscopy for approximately 200 galaxies in the near field.  The 
sample was analyzed with the FAST spectrograph operating on the Tillinghast 
\begin{figure}[t]
\begin{center}
\figurenum{4}
\includegraphics[width=0.9\textwidth]{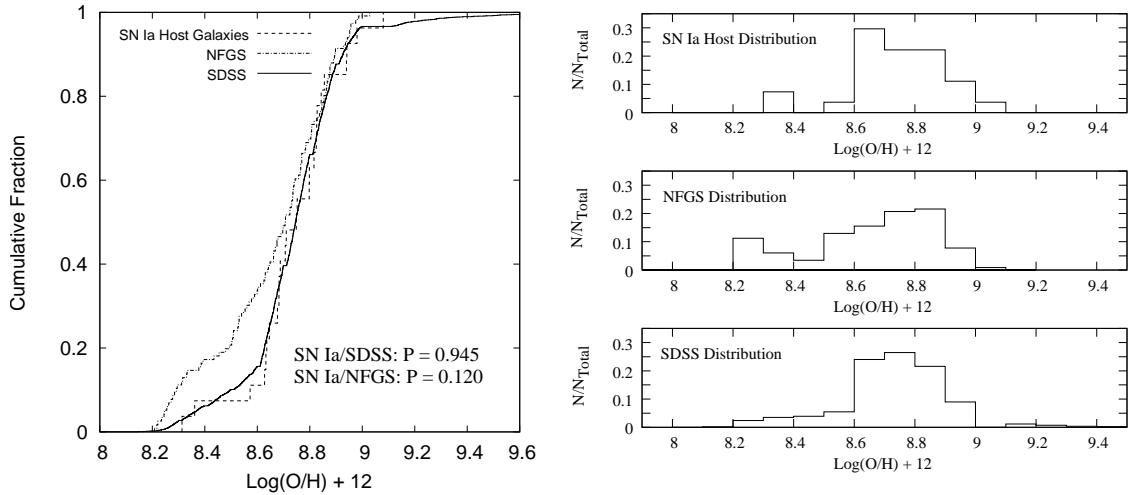}
\caption{Cumulative fraction plots for our SN host galaxy sample, 
NFGS galaxy sample, and the SDSS galaxy sample.  KS-test finds 
the probability of the host galaxy distribution and the NFGS 
distribution being drawn from the same distribution to be 12.0$\%$.  
Moreover, we find that the likelihood of the host galaxy sample and 
the SDSS sample being drawn from identical galaxy distributions to 
be 94.5$\%$. Fig.  5. -- The metallicity distributions for the SN host galaxy, 
NFGS, and SDSS samples.}
\end{center}
\end{figure}
telescope and includes galaxies of every morphological type covering 8 magnitudes 
in luminosity \citep{jansen}(Hereafter J00).  We were able to calculate metallicities for 
116 galaxies using the integrated emission line EW presented in J00.  The second 
comparison distribution is a set of approximately 9,000 SDSS galaxies whose 
spectroscopic line strength data was obtained through the Carnegie Mellon University/University of 
Pittsburgh's SDSS Value-Added Catalog (CMU-VAC). We were able to obtain 
3,133 galaxy metallicities from the SDSS sample using the EW line strengths 
obtained from the CMU-VAC.  For consistency, host galaxy metallicities used for 
comparison with the NFGS and the SDSS metallicity distributions were calculated 
from emission line EW ratios, and both the NFGS and SDSS samples were limited 
to emission line galaxies.

We performed a Kolmogorov-Smirnov test to determine the likelihood that our host 
galaxies were drawn from similar distributions as the NFGS sample and the 
SDSS sample.  Figure 4 and Figure 5 show the cumulative fraction plots (CFP) and 
the histograms for the metallicity distributions in each sample, respectively.  
We find from a KS test that the observed SN host galaxies could be drawn from the 
NFGS sample (12.0$\%$ probability) or from the SDSS sample (94.5$\%$).  The 
consistency between Ia host distribution, the NFGS distribution, and particularly the SDSS 
distributions implies a high probability that the SN hosts do not have a unique metallicity
signature as compared to a general sample of field galaxies, suggesting that the probability 
of a SN Ia occuring is not strongly dependent on the metallicity of the host galaxy in the 
range 0.05Z$_{\sun}$ $\la$ Z $\la$ 3.5Z$_{\sun}$.  

\subsubsection{Age}

In each of the Figures (2-5) the observed behaviors fall into one of two categories.  
They either show a negligible effect of metallicity on the luminosity of type Ia supernovae, 
or they show a trend that opposes current theoretical predictions.  In either case, the 
results suggest that metallicity is unlikely to be the primary contributor to the variability 
observed in the peak luminosities of type Ia supernovae.  Another property known to be 
correlated with host galaxy morphology is population age.  The population age, which we use
as an approximation of 
the progenitor age, may be able to explain the luminosity variations of SNe Ia and the 
correlation of these luminosities with host galaxy morphological type (Figure 1).  The progenitor
age is the amount of time between the birth of the progenitor and the time of the supernova event.
We can 
investigate the effect of progenitor age on the SN Ia luminosity distribution using a 
simple model inspired by \citet{Um99}.  This is 
shown in Figure 6.  We first adopted the single degenerate Chandrasekhar mass model for 
our SN progenitor system.  We then randomly selected pairs of stars, M$_{1}$ and M$_{2}$, from 
a distribution of stars consistent with the IMF in \citet{ktg93}.  A constraint is placed on 
the secondary mass requiring:
\begin{eqnarray}
(M_{2} - M_{WD_{2}}) + M_{WD_{1}} \geq 1.4M_{\sun}  
\end{eqnarray}
where M$_{WD_{1}}$ and M$_{WD_{2}}$ are the subsequent white dwarf masses corresponding to 
main sequence masses M$_{1}$ and M$_{2}$ \citep{Dom99}.
This ensured that the secondary possessed the necessary mass required for M$_{WD_{1}}$
to attain the Chandrasekhar limit through mass accretion.  To prevent the inclusion of stars that 
explode as core collapse SNe, we limited M$_{1}$ and M$_{2}$ to be $\leq$ 8M$_{\sun}$.  
Finally, we assumed the SNe explosion occurs soon after the main sequence lifetime of 
the secondary, thus setting the progenitor age, i.e the delay time, by the lifetime of the secondary.

Figure 6a shows the 
distribution of progenitor age as a function of the M$_{1}$.  For a given primary mass, there 
is an upper and lower limit placed on the mass of the secondary, and thus the lifetime of 
the secondary.  The lower limit on the progenitor mass arises from the requirement that the secondary 
\begin{figure}[h!]
\begin{center}
\figurenum{6}
\includegraphics[width=0.75\textwidth]{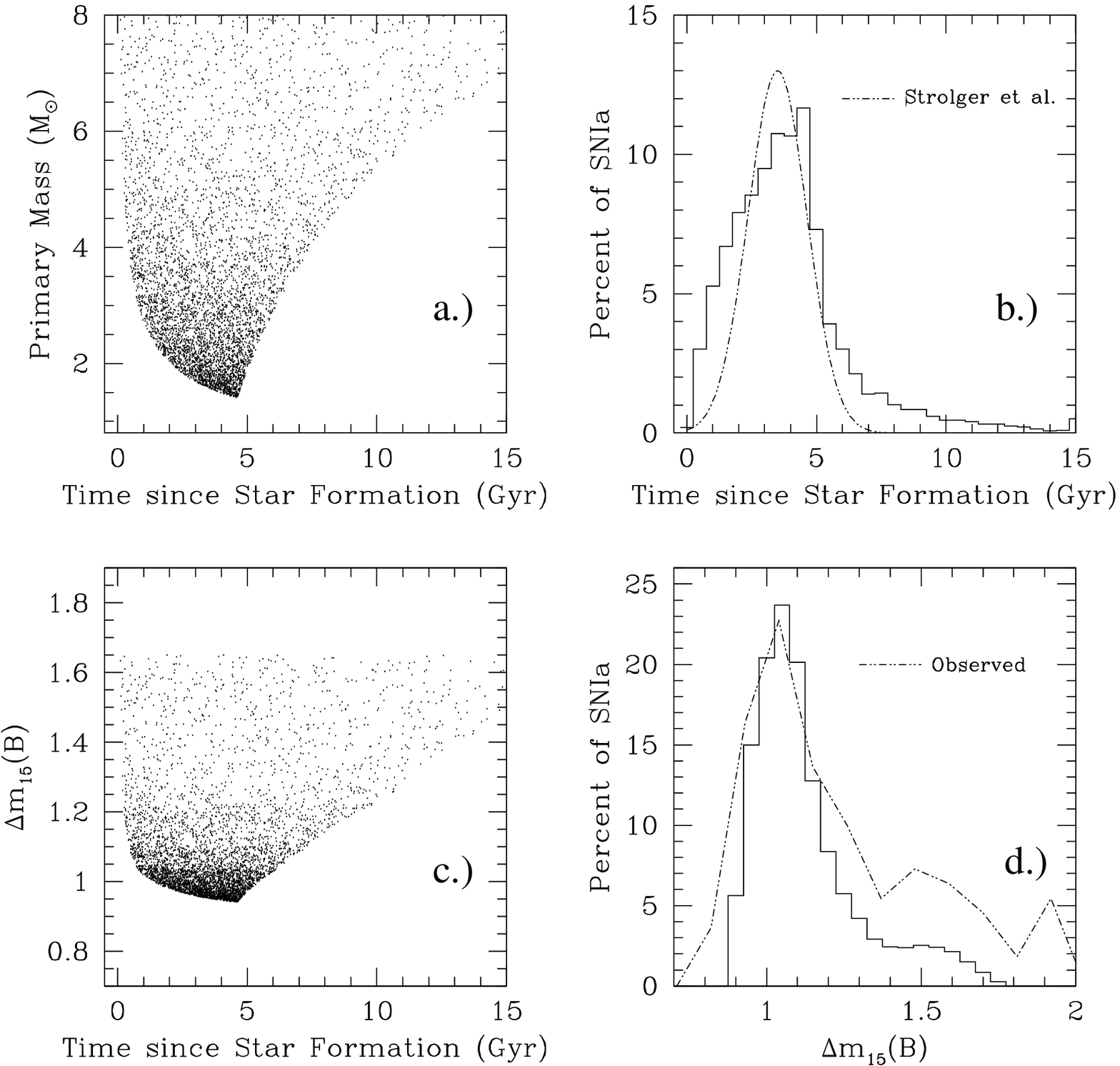}
\caption{The effect of progenitor age on the variations in type Ia 
supernova luminosity.  Progenitor age as a function of primary 
mass is given in (a) assuming an IMF as published in \citet{ktg93}.  
The distribution of progenitor age with the predictions of 
\citet{Strol03} over-plotted (b).  The SN Ia decline rate as a 
function of progenitor age (c), and the distribution of decline 
rates that results (d).}
\end{center} 
\end{figure}
star have enough mass to put the M$_{WD_{1}}$ over the Chandrasekhar limit.  On the other 
hand, the upper limit arises from a need for the secondary to have a lower mass than the 
primary.  Allowing the secondary to have a greater mass would simply exchange the 
respective labels of primary and secondary.  Given the nature of stellar evolution, an upper 
limit on mass becomes a lower limit on age, and vice versa.  The point where the two limits meet 
represents a system in which M$_{1}$ = M$_{2}$.  Binning the age axis in Figure 6a yields the age 
distribution of SNe Ia in Figure 6b.  We have over-plotted the derived distribution published 
in the results of \citet{Strol03} for the type Ia supernovae delay time, or the time between 
progenitor formation and the SN event.  Our simple model is in 
good agreement with an average type Ia supernova delay time around 3 Gyrs.  

Next, we set out to determine the expected effects that the progenitor age has on type Ia supernova 
decline rate by first approximating a linear fit to the plot of core mass versus main sequence mass
from \citet{Dom99} and converting our primary masses into white dwarf masses.  \citet{Um99} postulated 
that the M$_{^{56}Ni}$, and consequently the brightness of the SN Ia, increases as the C/O ratio of the 
progenitor increases.  Based on the postulate, they developed a model describing this dependency.  
Although the model should be treated with caution given that it is based on an unproven postulate, 
we use it in our toy model to merely provide a rough understanding on the effects of age on the decline rates of SNe Ia.  
Therefore, using these $^{56}$Ni yields and the M$_{^{56}Ni}$ to $\Delta$m$_{15}$(B) 
relations described in $\S$ 3.2.1, we converted the white dwarf masses to the expected SN decline rates.  The 
resultant decline rate versus age scatter plot is shown in Figure 6c.  By binning the $\Delta$m$_{15}$(B) 
\begin{figure}[t]
\begin{center}
\figurenum{7}
\includegraphics[width=0.7\textwidth]{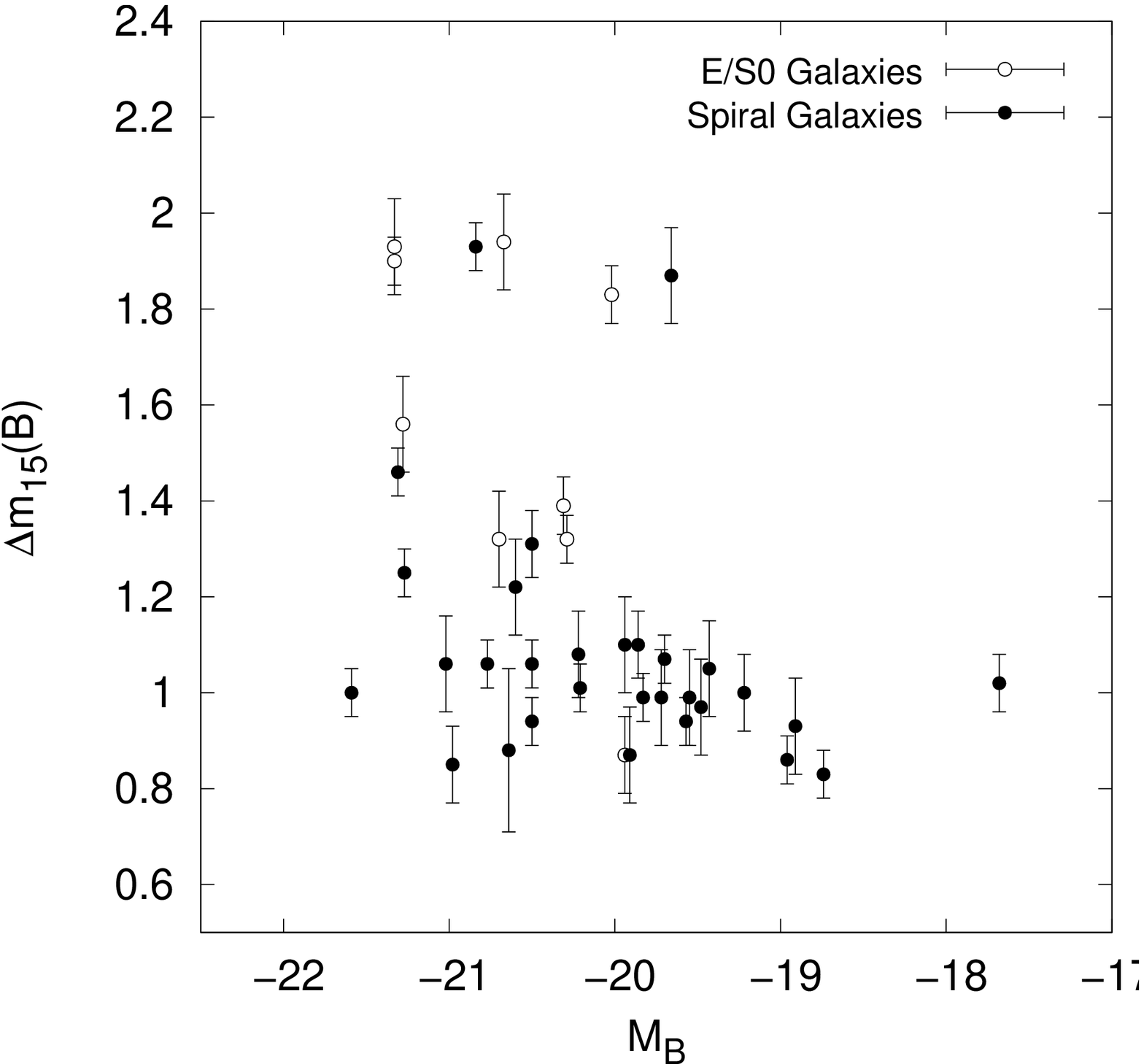}
\caption{Decline rate versus Host Galaxy Absolute magnitude.  
Distribution shows an absence of dimmer SNe Ia in the low 
host galaxy luminosity regime.}
\end{center}
\end{figure}
axis we obtain the expected decline rate distribution for type Ia supernovae in Figure 6d.  This 
figure shows that the age of the progenitor can result in a variation in decline rate similar to that 
which is observed for SNe Ia.  This consistency is compelling evidence for age, 
not metallicity, acting as the primary source of SN Ia diversity.  Future studies will obtain 
spectra of elliptical galaxies for use in absorption line metallicity and age estimations using 
the stellar population synthesis models of \citet{Worthey94}.  This will 
enable us to investigate the effects of age on the properties of type Ia supernovae directly.       

\subsubsection{Absolute B Magnitude}

H00 showed, for a sample of nearby SN Ia host galaxies, a trend (with high dispersion) 
indicating that the SN Ia decline rate increases, and consequently its maximum luminosity 
decreases, with increased host galaxy luminosity.  \citet{HW99} were able to 
show that the integrated luminosity of a galaxy is correlated with its global metallicity, as 
the brighter, more massive galaxies were able to better retain SN heavy metal ejecta.  
Therefore, H00 argued that any correlation observed between decline rate and absolute 
magnitude might manifest itself as a correlation between decline rate and metallicity.  
Figure 7 shows the absolute B magnitude of our host galaxy sample versus 
$\Delta$m$_{15}$(B).   Although Figure 7 does not show the gradual trend observed by 
H00, the plot does show less scatter for the least luminous galaxies likely due to
a combination of two effects.  The first being a selection effect brought about by fewer SNe 
occurring in smaller galaxies.  Such an effect would suggest that if we had more SNe from low 
luminosity galaxies, then the bias toward bright SNe Ia in the low galaxies luminosity regime would 
disappear.  However, according to Figure 1, fainter SNe are 
predominantly hosted by elliptical galaxies which are on average brighter than spiral galaxies.
Therefore, we see in Figure 7 a tendency for fainter SNe to be hosted by large, bright 
galaxies.  The combination of these two effects contribute to the distribution observed in Figure 7.

\begin{figure}[h]
\begin{center}
\figurenum{8}
\includegraphics[width=1.0\textwidth]{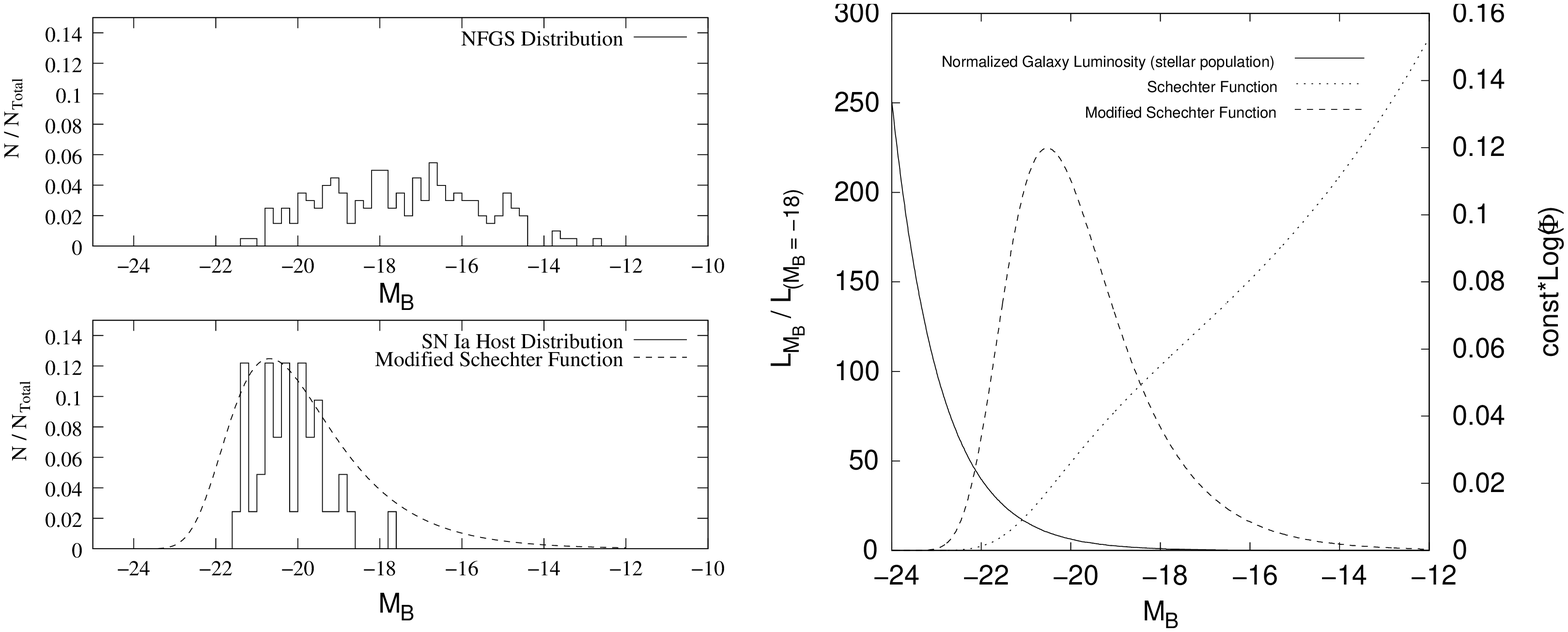}
\caption{Absolute magnitude distribution for the NFGS galaxy 
and Ia host galaxy samples.  Histogram reveals the host 
galaxy distribution to be statistically brighter than the NFGS 
galaxy sample.  Broken line is the theoretical SNe Ia
distribution as predicted by our modified Schechter Function.
Fig.  9. --  Modified Schechter Function:  Solid line shows the 
relative galactic luminosity, and consequently the galactic 
stellar population as a function of galaxy absolute B 
magnitude.  The dot-dot-dash line shows the Schechter 
function as described by Schechter(1976).  The modified 
Schechter function,dashed line, gives the relative probability 
of finding a type Ia SN as a function of galactic B filter magnitude. }
\end{center}
\end{figure}

Figure 8 shows the histograms of the NFGS and the SN Ia host galaxy distributions.  The 
two distributions are clearly dissimilar owing to the nearly 2.5 magnitude discrepancy 
observed between the average magnitudes of the respective galaxy samples.  The 
NFGS selected galaxies covering a wide range of luminosities (8 magnitudes), thus we 
would expect a broad distribution with little evidence of bias, as seen in Figure 8.  On 
the other hand, the SN Ia host galaxy selection process was not subject to such regulations.  
However, several selection effects were ultimately present in the SN Ia search.  
Firstly, the selection of target galaxies in the supernova Ia search often involved point 
searches that focused on well known, and inevitably more luminous targets that would 
bias the SN host galaxies toward the higher luminosity regime.  Furthermore, bias was 
introduced due to the non-uniformity of the Luminosity Function (LF) of galaxies.  The form 
of the LF known as the Schechter function \citep{Schec76},
\begin{eqnarray}
\Phi(M) \sim 10^{-0.4(\alpha+1)M}e^{-10^{0.4(M^{*}-M)}}  
\end{eqnarray}
with $\alpha$ = -1.17 and M$^{*}$ = -20.73 given by the Century Survey \citep{Gell97},
is plotted in Figure 9.

The LF illustrates that high luminosity galaxies are in the minority throughout the universe.  Consequently,
the probability of finding a SN Ia in a high luminosity galaxy is small, and the overall M$_{B}$ distribution 
of SN Ia host galaxies will be shifted to the lower magnitudes.  However, higher luminosity galaxies 
inherently have larger populations of stars than their lower luminosity
counterparts.  Figure 9 further reflects how the relative luminosity, and the stellar population of a galaxy changes with absolute B magnitude (function 
normalized to L$_{(M_{B} = -18)}$ = 1).  This curve assumes the luminosity in the B-band to be an adequate tracer of 
galactic mass.  Although not the best tracer \citep{mann05}, in the absence of good near-infrared H or K band 
measurements, B-band should be sufficient for this analysis.  We can see from the figure that although 
high luminosity galaxies are more rare than low 
luminosity galaxies, they possess more stars and thus have an increased probability of hosting a SN Ia.  We can
investigate the combined effects of these two phenomena through a Modified Schechter Function (MSF) 
represented by the product of these two functions governing the biases.  This MSF, Figure 9, represents an approximate
probability distribution governing the most likely absolute magnitudes for galaxies hosting SNe Ia.  The MSF does a 
reasonable job in predicting the absolute magnitude distribution range for our set of SN Ia host galaxies (Figure 8).    

\subsubsection{Scalo Birthrate Parameter}

H$\alpha$ equivalent width versus $\Delta$m$_{15}$(B) is given in Figure 10 while the 
H$\alpha$ luminosity is plotted against decline rate in Figure 11.  The galaxies with no 
discernible H$\alpha$ emission are shown with their upper limits.  They illustrate the 
propensity for the fastest declining type Ia SNe to occur in low emission galaxies.  This result 
suggests that the current star formation is a galactic property at least partly responsible for
the trend discovered by H00 and seen in Figure 1.  Moreover, Figure 
10 shows the H$\alpha$ distribution to be bimodal.  The presence of a gap around an 
equivalent width of 18 \AA\ suggests the possibility that there are two distinct populations of 
\begin{figure}[t]
\begin{center}
\figurenum{10}
\includegraphics[width=0.8\textwidth]{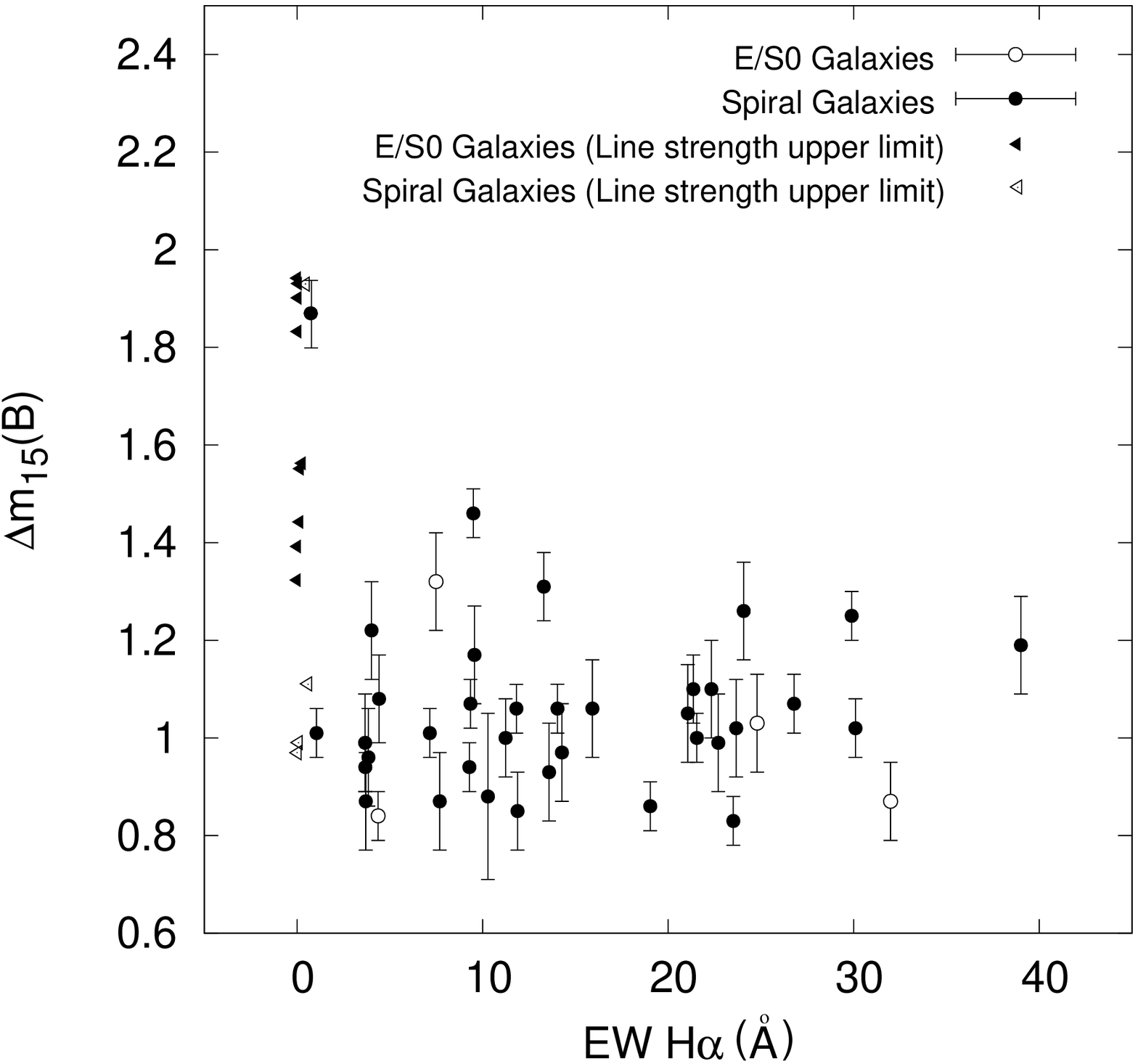}
\caption{Supernova decline rate versus H$\alpha$ equivalent width.  The gap present in the center 
of the distribution may suggest two distinct populations of type Ia supernovae.} 
\end{center}
\end{figure}
type Ia supernovae.  Scalo \textit{b} is plotted against decline rate in Figure 12.  Recalling the 
definition of Scalo \textit{b}, we know that current star formation relative to that of the past 
increases with increasing \textit{b}.  Such a gap might suggest 
the existence of one type of SN with a short delay time residing in high star forming galaxies, and another 
type having a longer delay time residing in low star forming galaxies.

Figure 13 shows the cumulative fraction plots for the three distributions of Scalo \textit{b}.  Our goal 
was to compare the inherent differences between the galaxies of these three surveys; therefore, it was important 
to minimize the sample differences brought about by selection effects present in the design of each survey.  
One such selection effect apparent in the NFGS was that the quoted line strengths 
reported by J00 did not include specified upper limits.  Such was not a reasonable expectation for
the automated line fitting procedures used by SDSS.  Consequently, weak line strengths that may not have been
recorded in the NFGS would have been recorded by the SDSS.  This would increase the relative 
populations of low star forming galaxies in the SDSS compared to the NFGS, and it would ultimately 
introduce uncertainty into our sample comparison.  Therefore, we imposed
a low end cutoff in the Ia host galaxy sample and the SDSS sample corresponding to the lowest Scalo \textit{b}
present in the NFGS.  Although this prevented our ability to test the distributions among low star formation 
galaxies, it did ensure that the three distributions could be accurately compared at high Scalo \textit{b} 
without the introduction of uncertainties due to inconsistent survey design at low Scalo \textit{b}.  
The result of the KS-test reveal that the probability of the host galaxy distribution being drawn from 
the same distributions as the SDSS and the NFGS sample to be 3.2$\%$ and 0.9$\%$, respectively.  
\begin{figure}[t]
\begin{center}
\figurenum{11}
\includegraphics[width=0.8\textwidth]{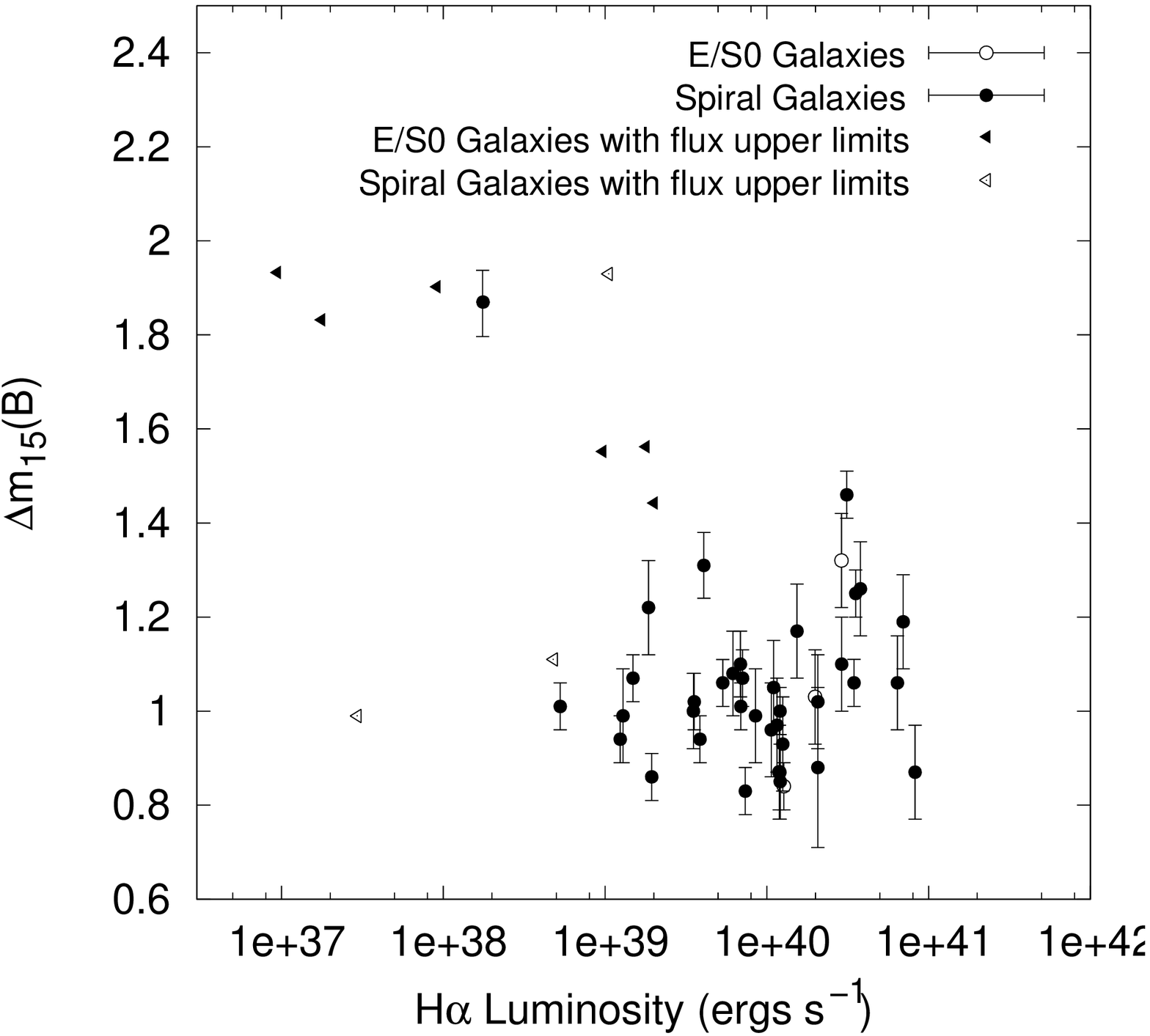}
\caption{H$\alpha$ Luminosity versus $\Delta$m$_{15}$(B)}
\end{center}
\end{figure}

Qualitatively comparing the NFGS and SNe Ia host galaxy sample lead to the following differences.
The aforementioned gap in H$\alpha$, and consequently Scalo \textit{b}, along with a relative lack of low 
SFR galaxies in the NFGS result in the the SN Ia cumulative fraction plot
increasing at a more rapid rate than the NFGS CFP.  The second, and more intriguing difference is the 
lack of high SFR 
galaxies in the SN Ia host galaxy sample.  Although similar at moderate Scalo \textit{b}, the same disparity is seen 
between the Ia host sample and the SDSS sample at high Scalo \textit{b}.  Both the NFGS and the SDSS 
distribution turn over around \textit{b} = 2.0, indicating the presence of high SFR galaxies in these respective 
samples.  It is for this reason that the KS-test yields such low probability for the null hypothesis that the Ia 
distributions was drawn from the SDSS sample. This is also apparent in Figure 14 showing the Scalo 
\textit{b} histograms for the three distributions.  

The similarities observed between the SDSS sample and the Ia host sample at moderate Scalo \textit{b}, 
combined with the intentionally non-partial nature of the NFGS suggest that the difference between the NFGS
and the Ia host galaxy sample at low relative SFR does not suggest a meaningful discrepancy between Ia hosts and the general 
galaxy population.  In reality, the difference is most likely a consequence of the NFGS selection process. 
KTC94 showed that Scalo \textit{b} is correlated with galaxy morphological type.  Consequently, we would 
expect a distribution of galaxies chosen to be uniform in morphological type and luminosity, such as the NFGS, 
to be likewise uniform in Scalo \textit{b}.  
\begin{figure}[t]
\begin{center}
\figurenum{12}
\includegraphics[width=0.8\textwidth]{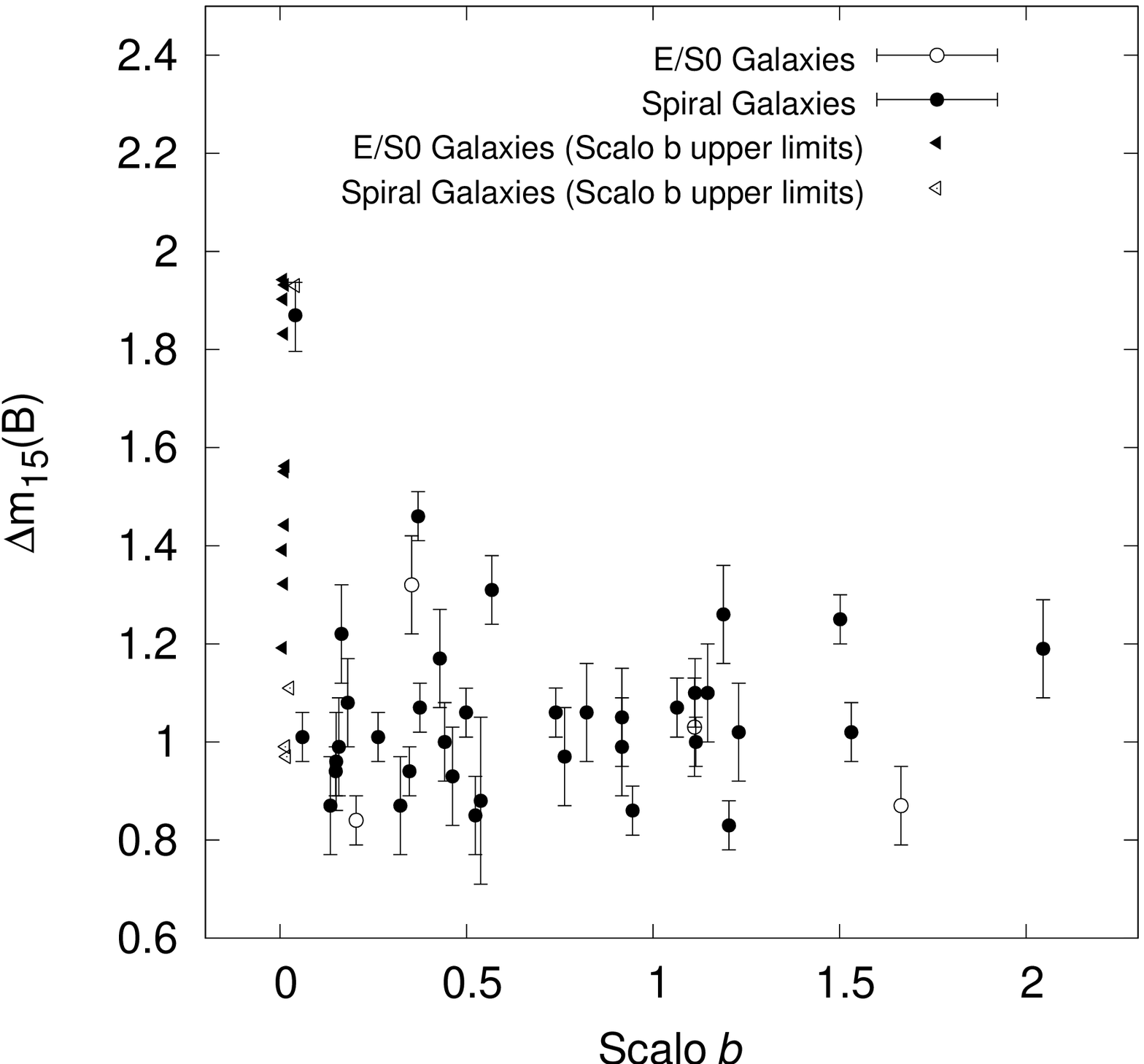}
\caption{Scalo \textit{b} parameter \citep{Sc86} versus SN Ia light curve decline rate.  Scalo \textit{b}
represents the current star formation relative to average past star formation.}
\end{center}
\end{figure}

However, the nature of the Ia host distribution at high Scalo \textit{b} is inconsistent with that observed in both 
the NFGS and the SDSS galaxies samples.  Galaxies with current star formation higher
than approximately 2 times the average past SF seem to be selected out of the Ia host galaxy sample.  A Monte 
Carlo test gives the probability of selecting 39 Scalo \textit{b} values at random from the SDSS sample and 
fortuitously obtaining all values below 2.05 to be $\sim$0.05$\%$.  Such a result implies a high significance 
for this cutoff.

A property of type Ia supernovae that has the potential to explain this rejection is the SN delay time, or 
alternately the time between progenitor formation and the supernova event.  We investigated the 
implications of this cutoff given the following assumptions:
1.)  The high star forming galaxies in the SN Ia sample have SFHs that are described by an 
exponential decline in SFR followed by a recent burst.  We assume that the progenitor for each SN Ia hosted 
by a high star formation galaxy was thus formed during the current burst.  2.)  There is a direct correlation between 
the number of SDSS galaxies observed in a particular Scalo \textit{b} range (Figure 14) and the duration over which the 
average galaxy spends producing stars at that star formation rate (t$_{burst}$), and 3.)  The ratio of the duration 
of a burst, with its peak at Scalo \textit{b} = \textit{b}$^{\prime}$, to the total lifetime of the galaxy ($\sim$ 14 Gyr) can be found 
through evaluation of the following: 
\begin{eqnarray}
Ratio & = & \frac{\int_{b'}^{\infty} t_{burst}(b)db}{\int_{0}^{\infty} t_{burst}(b)db}
\end{eqnarray} 
Given these three assumptions it is possible to place an approximate lower limit on the SN Ia delay time, $\tau$$_{min}$.  
\begin{figure}[t]
\begin{center}
\figurenum{13}
\includegraphics[width=1.0\textwidth]{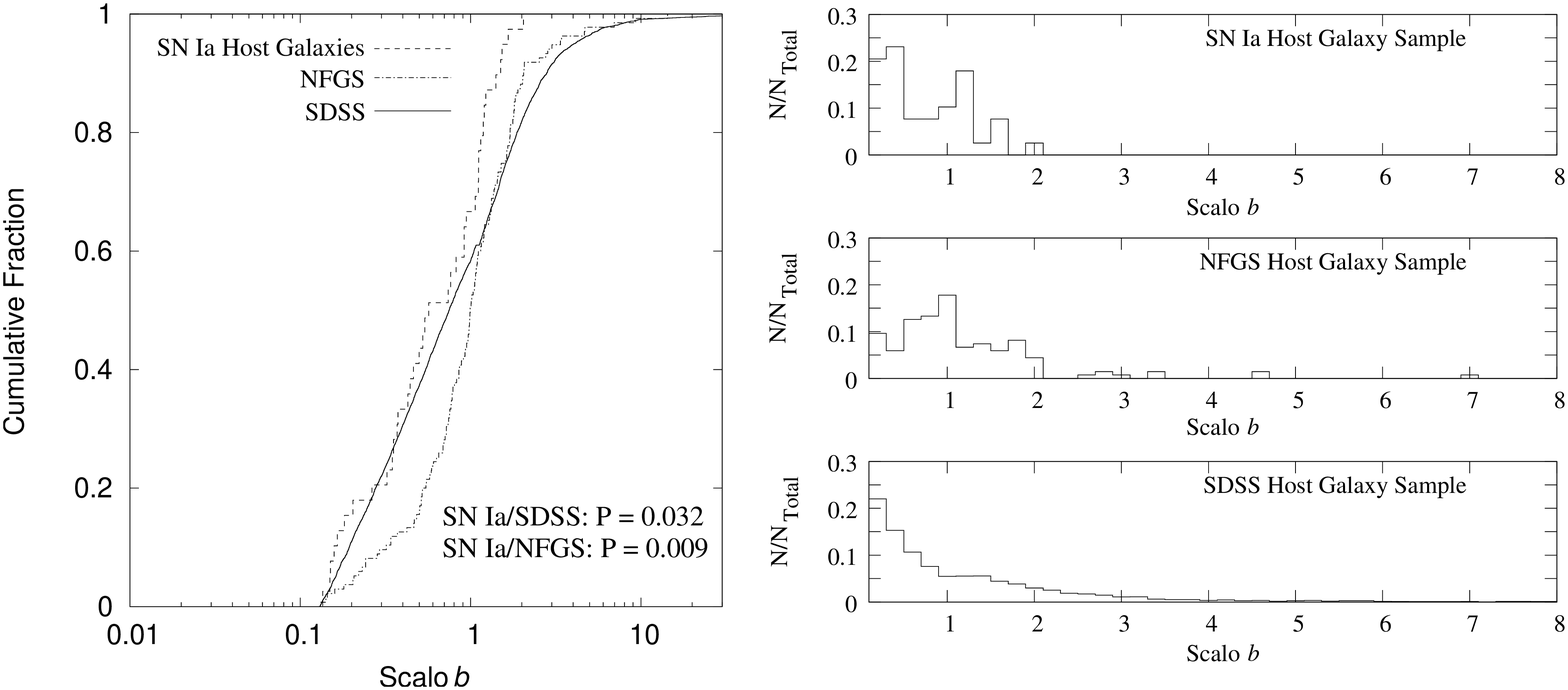}
\caption{Cumulative fraction plots for our type Ia supernovae host galaxy sample, 
the NFGS sample of galaxies, and the SDSS galaxy sample.  Fig. 14. ---  Scalo \textit{b} 
distribution in the SN Ia host sample, the NFGS sample, and the SDSS sample.
Note the cutoff in the SNe Ia host galaxy sample at \textit{b} $\sim$ 2.}
\end{center}
\end{figure}
Our observed cutoff might suggests that galaxies with current star formation rates higher than \textit{b} $\sim$ 2.0 are in the midst of a 
star formation burst that is too short to have created a SN Ia progenitor while allowing ample time for it to evolve and explode.  
By approximating an average galactic lifetime of 14 Gyrs, an evaluation of equ.(3) for \textit{b}$^{\prime}$ = 2.0 results in a  
$\tau$$_{min}$ $\sim$ 2.0 Gyrs.  Recent work conducted by \citet{GalYam} attempted to constrain $\tau$$_{min}$ by comparing 
the theory and observation of the SN Ia redshift distribution.  Beginning with theoretical functions governing the SFH and SN Ia delay time
they calculated the expected redshift distribution of SNe Ia and compared this to the observed distribution of supernovae discovered 
by the Supernova Cosmology Project.  The SFH function and SN Ia delay function are degenerate and thus could not be simultaneously 
constrained.  A SFH function had to be assumed in order to constrain the SN Ia delay function, and vice versa.  Assuming the SFH 
function predicted by \citet{madau98}, the results predict $\tau$$_{min}$ = 1.7\textit{h}$^{-1}$ Gyrs, $\sim$ 1.2 Gyrs, at the 95$\%$ 
confidence level.  However, they go on to show that a longer characteristic delay time lower limit, $\tau$$_{min}$ $\ge$ 
3\textit{h}$^{-1}$ Gyrs (2.1 Gyrs), is allowed if the SFH model of \citet{lanz02} is assumed.

\subsection{Hubble Residual}
Assuming a positive correlation between between galactic and progenitor 
metallicity, our results indicate that it is unlikely that variations in progenitor metallicity can entirely
account for the large brightness variations observed in SNe Ia at peak luminosity.  Although not a primary contributor, 
metallicity may contribute to more subtle variations at the level of the SN Ia intrinsic scatter.  
This can be tested by plotting the SN Hubble residuals, which are expected to have variations on 
the order of 0.18 mag \citep{Jha02}, versus galactic metallicity (Figure 15a).  The plot appears to 
show a slight trend for higher metallicity galaxies to produce SNe Ia with negative Hubble diagram residual. 
\begin{figure}[t]
\begin{center}
\figurenum{15}
\includegraphics[width=0.8\textwidth]{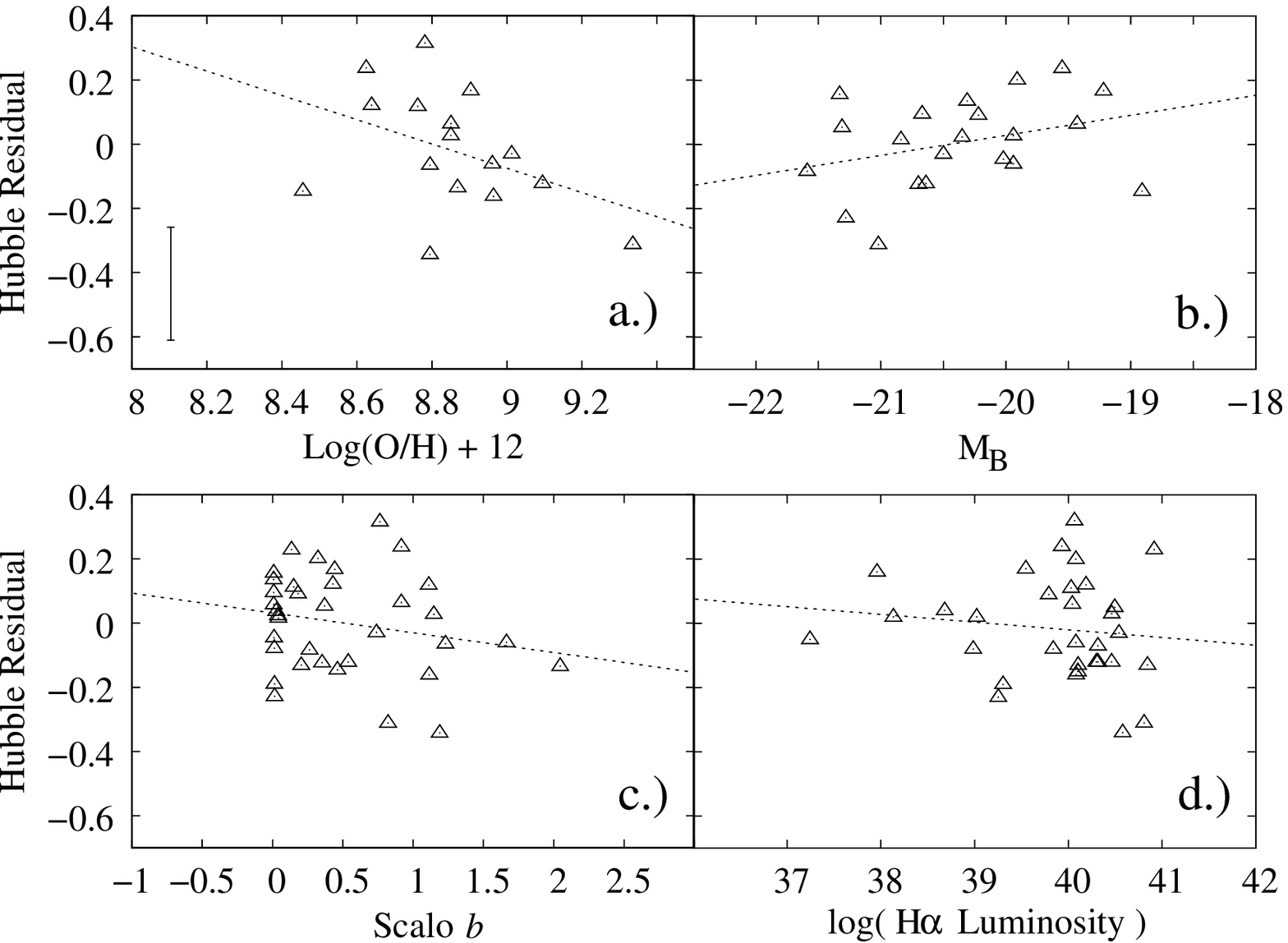}
\caption{Systematic errors present in the SN Hubble residuals.  Figure shows the relationship between Hubble residual 
and host galaxy metallicity (a), host galaxy absolute magnitude (b), Scalo \textit{b} (c), and the log of H$\alpha$ luminosity (d).  
A representative error bar is given in the lower left.  Monte Carlo tests reveal the correlations to be statistically insignificant 
to varying degrees.}    
\end{center}
\end{figure}
A Monte Carlo simulation reveals this correlation amplitude to be 90$\%$.  
The test was conducted as follows.  We generated 16 Gaussian distributed random numbers using a standard 
deviation of 0.18 and assigned to each one a metallicity measurement from our data.  We then plotted the 
metallicity versus this theoretical residual and calculated the slope of the best fit line.  Repeated trials 
allowed us to determined the likelihood of obtaining a best fit slope to the data greater than or equal to 
the absolute value of the slope observed in the best fit to Figure 15a.  This is a less than 2$\sigma$ detection and 
should be treated with caution.  Although, it could suggest metallicity to be a secondary parameter affecting the 
brightness of SNe Ia at the 10$\%$ level.  Nevertheless, continued study is need to bring about a more conclusive understanding.
Figures 15(a-c) show similar plots for absolute magnitude, Scalo \textit{b}, and the log of H$\alpha$ luminosity.  We found 
insignificant correlations for these three parameters with Monte Carlo simulations placing the linear fits near the 
75$\%$, 75$\%$, and 50$\%$ confidence levels, respectively.

\subsection{Summary}
We have analyzed the globally integrated spectra for a sample of type Ia supernova host galaxies in order 
to investigate the possible systematic effects that the host galaxy environment has on the properties
of type Ia supernovae.  We looked for direct correlations between the decline rates of type Ia supernovae
and host galaxy metallicity, absolute B filter magnitude, and star formation rate.  We further looked for 
correlations between these galactic parameters and the resident supernova's Hubble best fit residual.  Finally,
we investigate the systematic differences between SN Ia host galaxies and the general galactic population through a 
series of Kolmogorov-Smirnov tests.  The main results are as follows:

1. We find no correlation between spiral galaxy metallicity and SN Ia decline rate.  We find 
significant decline rate variability at fixed host galaxy metallicity 
implying a small impact of metallicity on the peak luminosity of type Ia supernovae.    
We find that SNe at smaller PGDs have a higher average luminosity than those residing further from their galaxy's nucleus.  As 
metallicity is predicted to decrease as a function of PGD in spiral galaxies, this result stands on contrast to the 
predictions made by combining the results of TBT03, \citet{Hof03}, and \citet{garn04}.  Furthermore, a KS-test 
shows the SN Ia host galaxy metallicity distribution to be statistically similar to the 
the NFGS, and particularly the SDSS metallicity distributions.  Our results also indicate that the progenitor age can have a 
significant impact on the variations in type Ia decline rates.  Assuming that the global galactic metallicity is approximately 
correlated with the progenitor metallicity, this implies that it is the age, and not the metallicity,  
of the progenitor that is the greater contributor to the inhomogeneities in type Ia supernovae explosions.  

2. The gradual trend found by H00 between host galaxy absolute magnitude 
and supernova decline rate is not seen in our galaxy sample.  Nevertheless, similar to the results of H00, we do find a lack 
of faint SNe in the low luminosity galaxy regime.  This is most likely a combination of the expected lack of SNe in small 
galaxies and the behavior observed in Figure 1 that shows the tendency for low luminosity SNe Ia to be 
hosted almost exclusively by early type galaxies.  Since ellipticals are on average brighter than other 
galaxies on the Hubble sequence, we would expect low luminosity SNe Ia to confine themselves to brighter 
host galaxies.  The average absolute B magnitude of our host galaxies are found to be systematically brighter than 
the galaxies of the NFGS.  The phenomenon can be attributed to the combined selection effect due 
to the increased population of stars in high luminosity galaxies and the non-uniformity of the luminosity 
function of galaxies.   

3. We do not see evidence for a dependency of supernova decline rate on Scalo \textit{b} in host galaxies with active
star formation.  However, we do see a discontinuity is the relationship between star formation rate and star formation history.
To the extent that all of the fast declining SNe Ia studied have been hosted by galaxies with low, to non-existent, star formation. 
Moreover, the distribution of Scalo \textit{b} for the host galaxy sample shows bimodal 
behavior suggesting the possibility of two distinct populations of type Ia supernovae 
possessing different delay times.  The host galaxy Scalo \textit{b} distribution also shows an unexpected cutoff at \textit{b} 
$\sim$ 2 that might be indicative of a finite lower limit on the delay time of type Ia supernovae.  We approximate 
this lower limit to be 2.0 Gyrs, slightly higher than the best value of 1.2 Gyrs obtained by \citet{GalYam}.  Future 
refinements to the SFH models should help to finally pin down the true characteristic delay time.  

4. Our tests to determine the effects of 
host galaxy environment on the SN Hubble residual all proved to be 
inconclusive to varying degrees.  However, a 90\% significance to a trend tying metallicity to Hubble residual, 
though requiring more study to prove or disprove, could suggest that metallicity, though not likely responsible for the diversity if SNe Ia 
brightnesses on the order of 1 magnitude, could be responsible for more subtle brightness variation seen 
at the 0.1 magnitude level.   A continuation of this work currently underway is attempting to increase the host 
galaxy population used in this study which will enable us to improve the likelihood of detecting statistically 
significant results. 

\acknowledgments
We would like to thank Lisa J. Kewley for supplying an IDL script used to perform host galaxy 
extinction corrections and metallicity determination.  Partial funding for this project was 
supplied by the Notre Dame Center for Applied Mathematics.  This research has made use of the NASA/IPAC 
Extragalactic Database (NED) which is operated by the Jet Propulsion Laboratory, California Institute of 
Technology, under contract with the National Aeronautics and Space Administration.

\clearpage
\begin{deluxetable}{lcccc}
\tabletypesize{\scriptsize}
\tablewidth{0pt}
\tablenum{1}
\tablecaption{The Host Galaxy Sample\label{tbl-1}}
\tablehead{
\colhead{Galaxy} & \colhead{SN} & \colhead{P.A.} & \colhead{Scan Width} & 
\colhead{Aperture Width}}
\startdata

NGC 4536 & 1981B & 30 & 120\arcsec & 51\arcsec \\
NGC 3627 & 1989B & 90 & 3\arcmin & 71\arcsec \\
NGC 4639 & 1990N & -10 & 1\arcmin & 51\arcsec \\
CGCG 111 016 & 1990O & -67 & 15\arcsec & 23\arcsec \\
NGC 4527 & 1991T & -25 & 120\arcsec & 46\arcsec \\
IC 4232 & 1991U & 5 & 20\arcsec & 31\arcsec \\
NGC 4374 & 1991bg & -50 & 3\arcmin & 38\arcsec \\
Anon & 1992J & 85 & 15\arcsec & 26\arcsec \\
IC 3690 & 1992P & -5 & 15\arcsec & 18\arcsec \\
Anon & 1992ag & 100 & 15\arcsec & 29\arcsec \\
Anon & 1992bp & 40 & 15\arcsec & 17\arcsec \\
CGCG 307 023 & 1993ac & 150 & 15\arcsec & 18\arcsec \\
NGC 4526 & 1994D & 20 & 90\arcsec & 46\arcsec \\
NGC 4493 & 1994M & -20 & 20\arcsec & 34\arcsec \\
CGCG 224 104 & 1994Q & 70 & 10\arcsec & 14\arcsec \\
NGC 4495 & 1994S & -45 & 20\arcsec & 46\arcsec \\
NGC 3370 & 1994ae & -35 & 1\arcmin & 23\arcsec \\
NGC 2962 & 1995D & -5 & 1\arcmin & 46\arcsec \\
NGC 2441 & 1995E & -20 & 1\arcmin & 51\arcsec \\
IC 1844 & 1995ak & 100 & 15\arcsec & 36\arcsec \\
NGC 3021 & 1995al & 100 & 30\arcsec & 46\arcsec \\
UGC 03151 & 1995bd & 100 & 15\arcsec & 51\arcsec \\
Anon & 1996C & 0 & 15\arcsec & 23\arcsec \\
NGC 2935 & 1996Z & -30 & 2\arcmin & 49\arcsec \\
Anon & 1996ab & -30 & 0\arcsec & 15\arcsec \\
NGC 5005 & 1996ai & -25 & 2\arcmin & 51\arcsec \\
Anon & 1996bl & 80 & 15\arcsec & 17\arcsec \\
NGC 0673 & 1996bo & 0 & 45\arcsec & 54\arcsec \\
UGC 03432 & 1996bv & 130 & 15\arcsec & 46\arcsec \\
NGC 2258 & 1997E & 150 & 15\arcsec & 29\arcsec \\
NGC 4680 & 1997bp & 30 & 20\arcsec & 40\arcsec \\
NGC 3147 & 1997bq & 0 & 90\arcsec & 68\arcsec \\
Anon & 1997br & -20 & 20\arcsec & 63\arcsec \\
NGC 5490 & 1997cn & 10 & 20\arcsec & 34\arcsec \\
NGC 0105 & 1997cw & 0 & 30\arcsec & 46\arcsec \\
UGC 03845 & 1997do & -10 & 30\arcsec & 46\arcsec \\
NGC 5440 & 1998D & 50 & 30\arcsec & 23\arcsec \\
NGC 6627 & 1998V & 80 & 15\arcsec & 34\arcsec \\
NGC 4704 & 1998ab & 45 & 30\arcsec & 29\arcsec \\
NGC 3982 & 1998aq & 90 & 1\arcmin & 57\arcsec \\
NGC 6495 & 1998bp & 20 & 20\arcsec & 34\arcsec \\
NGC 3368 & 1998bu & 45 & 2\arcmin & 57\arcsec \\
NGC 0252 & 1998de & 90 & 30\arcsec & 40\arcsec \\
CGCG 302 013  & 1998di & $-$30 & 0\arcsec & 17\arcsec \\
UGC 00139 & 1998dk & 80 & 15\arcsec & 51\arcsec \\
Anon & 1998dm & 20 & 15\arcsec & 68\arcsec \\
Anon & 1998dx & -30 & 10\arcsec & 29\arcsec \\
UGC 03576 & 1998ec & 130 & 30\arcsec & 38\arcsec \\
UGC 00646 & 1998ef & 100 & 15\arcsec & 19\arcsec \\
NGC 0632 & 1998es & $-$20 & \nodata & 31\arcsec \\
CGCG 180 022 & 1999X & -20 & 15\arcsec & 17\arcsec \\
NGC 2595 & 1999aa & -20 & 1\arcmin & 57\arcsec \\
NGC 6063 & 1999ac & -30 & 30\arcsec & 34\arcsec \\
NGC 2841 & 1999by & 65 & 2\arcmin & 51\arcsec \\
NGC 6038 & 1999cc & 90 & 30\arcsec & 50\arcsec \\
NGC 6411 & 1999da & 70 & 20\arcsec & 31\arcsec \\
NGC 6951 & 2000e & 0 & \nodata & 32\arcsec \\
\enddata
\end{deluxetable} 

\clearpage
\begin{deluxetable}{lccccccccccc}
\tablewidth{0pt}
\tablenum{2}
\rotate
\tabletypesize{\scriptsize}
\tablecaption{Emission Line Equivalent Widths (\AA)\label{tbl-2}}
\tablehead{
\colhead{Galaxy} & \colhead{SN} & \colhead{[OII]$\lambda$3727} & \colhead{H$\beta$$\lambda$4861} &
\colhead{[OIII]$\lambda$4959} & \colhead{[OIII]$\lambda$5007} & \colhead{[OI]$\lambda$6300} &
\colhead{[NII]$\lambda$6548} & \colhead{H$\alpha$$\lambda$6562} &
\colhead{[NII]$\lambda$6584} & \colhead{[SII]$\lambda$6717} & \colhead{[SII]$\lambda$6731}}
\startdata
NGC 4536 & 1981B & 10.95 & 2.77 & 0.65 & 2.29 & 1.03 & 2.91 & 21.36 & 9.18 & 5.29 & 3.93 \\
NGC 3627 & 1989B & 1.27 & 1.74 & 0.10 & 0.40 & 0.34 & 1.73 & 13.30 & 6.27 & 2.46 & 2.06 \\
NGC 4639 & 1990N & 20.76 & 1.39 & 0.29 & 1.65 & 0.88 & 1.14 & 9.34 & 5.23 & 3.00 & 1.76 \\
CGCG 111 016 & 1990O & 4.84 & 0.00 & 0.00 & 0.13 & 0.49 & 0.41 & 3.85 & 3.56 & 0.87 & 1.45 \\
NGC 4527 & 1991T & 3.20 & 0.60 & 0.18 & 1.78 & 0.00 & 0.85 & 9.29 & 4.70 & 1.62 & 1.66 \\
IC 4232 & 1991U & 10.84 & 2.83 & 0.77 & 0.66 & 0.55 & 4.27 & 15.92 & 7.65 & 2.38 & 7.41 \\
NGC 4374 & 1991bg & \nodata & \nodata & \nodata & \nodata & \nodata & 0.00 & 0.04 & 0.44 & \nodata & \nodata \\
Anon & 1992J & \nodata & \nodata & \nodata & \nodata & \nodata & 0.00 & 0.25 & 0.39 & \nodata & \nodata \\
IC 3690 & 1992P & 13.08 & 0.42 & 0.36 & 1.13 & 1.06 & 1.17 & 7.69 & 5.20 & 2.45 & 2.37 \\
Anon & 1992ag & 42.14 & 9.17 & 3.87 & 13.52 & 0.99 & 7.30 & 39.01 & 12.95 & 12.98 & 6.03 \\
Anon & 1992bp & 8.73 & 0.00 & 0.00 & 3.04 & 0.00 & 1.89 & 7.49 & 5.64 & 5.18 & 6.12 \\
CGCG 307 023 & 1993ac & \nodata & \nodata & \nodata & \nodata & \nodata & \nodata & \nodata & \nodata & \nodata & \nodata \\
NGC 4526 & 1994D & \nodata & \nodata & \nodata & \nodata & \nodata & 0.05 & 0.00 & 0.16 & \nodata & \nodata \\
NGC 4493 & 1994M & \nodata & \nodata & \nodata & \nodata & \nodata & 0.18 & 0.12 & 0.19 & \nodata & \nodata \\
CGCG 224 104 & 1994Q & 25.97 & 5.60 & 0.37 & 2.15 & 0.82 & 2.00 & 24.80 & 6.61 & 6.10 & 3.85 \\
NGC 4495 & 1994S & 20.37 & 4.37 & 1.22 & 3.93 & 0.73 & 2.09 & 22.33 & 10.01 & 4.42 & 5.88 \\
NGC 3370 & 1994ae & 16.04 & 3.59 & 0.66 & 0.72 & 0.00 & 2.20 & 19.04 & 7.28 & 4.66 & 3.06 \\
NGC 2962 & 1995D & \nodata & \nodata & \nodata & \nodata & \nodata & 0.13 & 0.08 & 0.40 & \nodata & \nodata \\
NGC 2441 & 1995E & 12.06 & 2.55 & 0.40 & 1.42 & 0.13 & 1.58 & 11.82 & 5.99 & 3.51 & 2.62 \\
IC 1844 & 1995ak & 17.56 & 3.98 & 0.43 & 1.81 & 0.60 & 2.39 & 24.07 & 9.18 & 6.57 & 4.42 \\
NGC 3021 & 1995al & 13.29 & 4.35 & 0.49 & 1.49 & 0.76 & 2.87 & 23.51 & 9.69 & 4.01 & 3.51 \\
UGC 03151 & 1995bd & \nodata & \nodata & \nodata & \nodata & \nodata & 0.78 & 4.36 & 4.78 & \nodata & \nodata \\
Anon & 1996C & 17.18 & 2.37 & 0.00 & 1.46 & 0.80 & 4.38 & 14.27 & 7.75 & 5.54 & 3.57 \\
NGC 2935 & 1996Z & \nodata & \nodata & \nodata & \nodata & \nodata & 0.98 & 4.01 & 3.42 & \nodata & \nodata \\
Anon & 1996ab & \nodata & \nodata & \nodata & \nodata & \nodata & 0.61 & 3.71 & 2.87 & \nodata & \nodata \\
NGC 5005 & 1996ai & \nodata & \nodata & \nodata & \nodata & \nodata & 0.89 & 3.66 & 3.57 & \nodata & \nodata \\
Anon & 1996bl & 3.66 & 0.48 & 0.30 & 0.85 & 0.84 & 2.25 & 9.56 & 5.53 & 3.24 & 2.27 \\
NGC 0673 & 1996bo & 20.42 & 6.34 & 1.98 & 3.52 & 0.59 & 3.40 & 29.89 & 11.20 & 6.30 & 5.31 \\
UGC 03432 & 1996bv & 16.45 & 2.40 & 0.83 & 4.47 & 0.83 & 1.39 & 13.58 & 3.37 & 4.88 & 4.30 \\
NGC 2258 & 1997E & \nodata & \nodata & \nodata & \nodata & \nodata & \nodata & \nodata & \nodata & \nodata & \nodata \\
NGC 4680 & 1997bp & 7.13 & 3.81 & 0.10 & 1.48 & 0.21 & 2.52 & 21.55 & 9.43 & 4.37 & 3.32 \\
NGC 3147 & 1997bq & \nodata & \nodata & \nodata & \nodata & \nodata & 0.83 & 7.15 & 4.08 & \nodata & \nodata \\
Anon & 1997br & 41.89 & 5.65 & 2.27 & 13.02 & 1.70 & 3.55 & 30.09 & 9.74 & 4.11 & 3.89 \\
NGC 5490 & 1997cn & \nodata & \nodata & \nodata & \nodata & \nodata & 0.03 & 0.03 & 0.03 & \nodata & \nodata \\
NGC 0105 & 1997cw & 17.21 & 3.79 & 2.11 & 5.67 & 0.75 & 3.74 & 23.67 & 9.71 & 5.15 & 4.08 \\
UGC 03845 & 1997do & 31.51 & 4.81 & 1.03 & 5.44 & 0.53 & 1.36 & 22.70 & 6.04 & 6.00 & 4.71 \\
NGC 5440 & 1998D & \nodata & \nodata & \nodata & \nodata & \nodata & 0.32 & 0.11 & 1.93 & \nodata & \nodata \\
NGC 6627 & 1998V & 7.25 & 2.58 & 0.20 & 3.86 & 0.42 & 2.73 & 14.03 & 9.03 & 2.85 & 2.20 \\
NGC 4704 & 1998ab & 11.00 & 2.93 & 0.23 & 0.70 & 0.27 & 2.67 & 10.29 & 7.54 & 4.98 & 1.38 \\
NGC 3982 & 1998aq & 17.40 & 5.47 & 0.91 & 3.32 & 0.65 & 3.30 & 27.44 & 11.15 & 5.81 & 3.98 \\
NGC 6495 & 1998bp & \nodata & \nodata & \nodata & \nodata & \nodata & 0.09 & 0.02 & 0.15 & \nodata & \nodata \\
NGC 3368 & 1998bu & \nodata & \nodata & \nodata & \nodata & \nodata & 0.64 & 1.04 & 1.92 & \nodata & \nodata \\
NGC 0252 & 1998de & \nodata & \nodata & \nodata & \nodata & \nodata & 0.28 & 0.45 & 1.79 & \nodata & \nodata \\
CGCG 302 013  & 1998di & \nodata & \nodata & \nodata & \nodata & \nodata & 0.31 & 0.31 & 0.31 & \nodata & \nodata \\
UGC 00139 & 1998dk & 13.07 & 4.59 & 0.62 & 2.48 & 0.67 & 1.86 & 21.06 & 7.18 & 5.45 & 3.91 \\
UGCA 017 & 1998dm & 30.67 & 5.43 & 2.42 & 9.19 & 0.83 & 1.38 & 26.79 & 5.32 & 6.92 & 4.60 \\
UGC 11149 & 1998dx & \nodata & \nodata & \nodata & \nodata & \nodata & 0.14 & 0.14 & 0.14 & \nodata & \nodata \\
UGC 03576 & 1998ec & 10.80 & 0.00 & 0.08 & 0.14 & 0.50 & 1.00 & 4.41 & 3.66 & 1.64 & 4.20 \\
UGC 00646 & 1998ef & \nodata & \nodata & \nodata & \nodata & \nodata & 0.08 & 0.00 & 0.87 & \nodata & \nodata \\
NGC 0632 & 1998es & 21.55 & 6.82 & 0.74 & 3.03 & 1.60 & 3.89 & 31.99 & 13.11 & 6.87 & 5.38 \\
CGCG 180 022 & 1999X & \nodata & \nodata & \nodata & \nodata & \nodata & 0.25 & 0.60 & 0.70 & \nodata & \nodata \\
NGC 2595 & 1999aa & 15.35 & 0.54 & 0.26 & 0.42 & 0.00 & 1.55 & 11.88 & 6.66 & 3.17 & 2.91 \\
NGC 6063 & 1999ac & 9.16 & 2.56 & 0.00 & 0.78 & 0.46 & 1.27 & 11.24 & 5.22 & 2.71 & 4.22 \\
NGC 2841 & 1999by & \nodata & \nodata & \nodata & \nodata & \nodata & 0.11 & 0.74 & 1.65 & \nodata & \nodata \\
NGC 6038 & 1999cc & \nodata & \nodata & \nodata & \nodata & \nodata & 1.53 & 9.50 & 4.54 & \nodata & \nodata \\
NGC 6411 & 1999da & \nodata & \nodata & \nodata & \nodata & \nodata & 0.02 & 0.00 & 0.06 & \nodata & \nodata \\
NGC 6951 & 2000E & \nodata & \nodata & \nodata & \nodata & \nodata & 1.26 & 3.68 & 2.84 & \nodata & \nodata \\
\enddata
\end{deluxetable} 

\clearpage
\begin{deluxetable}{lccccccccccc}
\tablewidth{0pt}
\tablenum{3}
\rotate
\tabletypesize{\scriptsize}
\tablecaption{Emission Line Fluxes (10$^{14}$ ergs cm$^{-2}$ s$^{-1}$) \label{tbl-3}}
\tablehead{
\colhead{Galaxy} & \colhead{SN} & \colhead{[OII]$\lambda$3727} & \colhead{H$\beta$$\lambda$4861} &
\colhead{[OIII]$\lambda$4959} & \colhead{[OIII]$\lambda$5007} & \colhead{[OI]$\lambda$6300} &
\colhead{[NII]$\lambda$6548} & \colhead{H$\alpha$$\lambda$6562} &
\colhead{[NII]$\lambda$6584} & \colhead{[SII]$\lambda$6717} & \colhead{[SII]$\lambda$6731}}
\startdata
NGC 4536 & 1981B & 3.11 & 0.76 & 0.20 & 0.68 & 0.30 & 0.86 & 6.32 & 2.72 & 1.58 & 1.18 \\
NGC 3627 & 1989B & 1.22 & 2.14 & 0.14 & 0.52 & 0.38 & 1.97 & 15.10 & 7.14 & 2.81 & 2.35 \\
NGC 4639 & 1990N & 4.94 & 0.60 & 0.13 & 0.71 & 0.34 & 0.45 & 3.68 & 2.05 & 1.19 & 0.70 \\
CGCG 111 016 & 1990O & 0.37 & 0.00 & 0.00 & 0.02 & 0.06 & 0.06 & 0.52 & 0.47 & 0.11 & 0.18 \\
NGC 4527 & 1991T & 0.84 & 0.38 & 0.09 & 0.59 & 0.00 & 0.35 & 3.76 & 1.90 & 0.64 & 0.65 \\
IC 4232 & 1991U & 0.10 & 0.60 & 0.13 & 0.13 & 0.08 & 0.72 & 2.68 & 1.29 & 0.40 & 1.24 \\
NGC 4374 & 1991bg & \nodata & \nodata & \nodata & \nodata & \nodata & \nodata & 0.03 & 0.32 & \nodata & \nodata \\
Anon & 1992J & \nodata & \nodata & \nodata & \nodata & \nodata & \nodata & 0.04 & 0.07 & \nodata & \nodata \\
IC 3690 & 1992P & 0.59 & 0.04 & 0.04 & 0.12 & 0.10 & 0.12 & 0.79 & 0.53 & 0.24 & 0.24 \\
Anon & 1992ag & 2.98 & 1.14 & 0.54 & 1.71 & 0.11 & 0.87 & 4.70 & 1.60 & 1.64 & 0.74 \\
Anon & 1992bp & 0.18 & 0.00 & 0.00 & 0.09 & 0.00 & 0.05 & 0.20 & 0.15 & 0.13 & 0.15 \\
CGCG 307 023 & 1993ac & \nodata & \nodata & \nodata & \nodata & \nodata & \nodata & \nodata & \nodata & \nodata & \nodata \\
NGC 4526 & 1994D & \nodata & \nodata & \nodata & \nodata & \nodata & 0.06 & \nodata & 0.20 & \nodata & \nodata \\
NGC 4493 & 1994M & \nodata & \nodata & \nodata & \nodata & \nodata & 0.03 & 0.16 & 0.03 & \nodata & \nodata \\
CGCG 224 104 & 1994Q & 0.84 & 0.27 & 0.02 & 0.12 & 0.04 & 0.09 & 1.09 & 0.29 & 0.27 & 0.17 \\
NGC 4495 & 1994S & 2.90 & 0.97 & 0.31 & 0.92 & 0.18 & 0.49 & 5.25 & 2.35 & 1.03 & 1.37 \\
NGC 3370 & 1994ae & 2.22 & 0.65 & 0.21 & 0.14 & 0.00 & 0.16 & 3.19 & 1.21 & 0.75 & 0.49 \\
NGC 2962 & 1995D & \nodata & \nodata & \nodata & \nodata & \nodata & 0.04 & 0.02 & 0.13 & \nodata & \nodata \\
NGC 2441 & 1995E & 1.47 & 0.44 & 0.08 & 0.26 & 0.02 & 0.25 & 1.88 & 0.95 & 0.54 & 0.40 \\
IC 1844 & 1995ak & 1.91 & 0.64 & 0.08 & 0.31 & 0.09 & 0.37 & 3.69 & 1.41 & 0.98 & 0.66 \\
NGC 3021 & 1995al & 3.79 & 1.93 & 0.24 & 0.69 & 0.31 & 1.19 & 9.76 & 4.02 & 1.64 & 1.44 \\
UGC 03151 & 1995bd & \nodata & \nodata & \nodata & \nodata & \nodata & 0.47 & 2.59 & 2.80 & \nodata & \nodata \\
Anon & 1996C & 0.57 & 0.13 & 0.00 & 0.08 & 0.04 & 0.22 & 0.70 & 0.37 & 0.27 & 0.17 \\
NGC 2935 & 1996Z & \nodata & \nodata & \nodata & \nodata & \nodata & 0.29 & 1.17 & 0.98 & \nodata & \nodata \\
Anon & 1996ab & \nodata & \nodata & \nodata & \nodata & \nodata & 0.04 & 0.22 & 0.16 & \nodata & \nodata \\
NGC 5005 & 1996ai & \nodata & \nodata & \nodata & \nodata & \nodata & 0.89 & 3.67 & 3.58 & 1.71 & 0.90 \\
Anon & 1996bl & 0.16 & 0.03 & 0.02 & 0.06 & 0.05 & 0.14 & 0.58 & 0.34 & 0.18 & 0.13 \\
NGC 0673 & 1996bo & 3.93 & 1.46 & 0.51 & 0.89 & 0.12 & 0.70 & 6.13 & 2.29 & 1.26 & 1.06 \\
UGC 03432 & 1996bv & 2.32 & 0.40 & 0.16 & 0.75 & 0.12 & 0.21 & 2.11 & 0.53 & 0.73 & 0.65 \\
NGC 2258 & 1997E & \nodata & \nodata & \nodata & \nodata & \nodata & \nodata & \nodata & \nodata & \nodata & \nodata \\
NGC 4680 & 1997bp & 1.92 & 1.15 & 0.03 & 0.47 & 0.06 & 0.75 & 6.42 & 2.81 & 1.30 & 0.99 \\
NGC 3147 & 1997bq & \nodata & \nodata & \nodata & \nodata & \nodata & 0.43 & 3.66 & 2.07 & \nodata & \nodata \\
Anon & 1997br & 3.13 & 0.89 & 0.48 & 2.24 & 0.24 & 0.41 & 3.51 & 1.15 & 0.62 & 0.60 \\
NGC 5490 & 1997cn & \nodata & \nodata & \nodata & \nodata & \nodata & 0.01 & 0.01 & 0.01 & \nodata & \nodata \\
NGC 0105 & 1997cw & 2.52 & 0.72 & 0.43 & 1.09 & 0.12 & 0.60 & 3.82 & 1.56 & 0.81 & 0.63 \\
UGC 03845 & 1997do & 4.79 & 0.93 & 0.23 & 1.11 & 0.09 & 0.22 & 3.68 & 0.98 & 0.95 & 0.74 \\
NGC 5440 & 1998D & \nodata & \nodata & \nodata & \nodata & \nodata & 0.11 & 0.04 & 0.64 & \nodata & \nodata \\
NGC 6627 & 1998V & 1.65 & 0.92 & 0.08 & 1.46 & 0.16 & 1.07 & 5.51 & 3.55 & 1.12 & 0.87 \\
NGC 4704 & 1998ab & 0.62 & 0.34 & 0.03 & 0.09 & 0.03 & 0.32 & 1.23 & 0.88 & 0.62 & 0.17 \\
NGC 3982 & 1998aq & 5.93 & 2.53 & 0.46 & 1.64 & 0.26 & 1.33 & 11.10 & 4.46 & 2.29 & 1.57 \\
NGC 6495 & 1998bp & \nodata & \nodata & \nodata & \nodata & \nodata & 0.04 & 0.01 & 0.06 & \nodata & \nodata \\
NGC 3368 & 1998bu & \nodata & \nodata & \nodata & \nodata & \nodata & 0.84 & 1.37 & 2.52 & \nodata & \nodata \\
NGC 0252 & 1998de & \nodata & \nodata & \nodata & \nodata & \nodata & 0.13 & 0.20 & 0.80 & \nodata & \nodata \\
CGCG 302 013  & 1998di & \nodata & \nodata & \nodata & \nodata & \nodata & 0.01 & 0.01 & 0.01 & \nodata & \nodata \\
UGC 00139 & 1998dk & 2.34 & 0.85 & 0.21 & 0.50 & 0.17 & 0.32 & 3.59 & 1.22 & 0.92 & 0.66 \\
UGCA 017 & 1998dm & 9.28 & 1.89 & 0.97 & 3.46 & 0.25 & 0.41 & 7.84 & 1.55 & 1.98 & 1.31 \\
UGC 11149 & 1998dx & \nodata & \nodata & \nodata & \nodata & \nodata & 0.01 & 0.01 & 0.01 & \nodata & \nodata \\
UGC 03576 & 1998ec & 0.89 & 0.00 & 0.01 & 0.02 & 0.08 & 0.16 & 0.71 & 0.59 & 0.26 & 0.64 \\
UGC 00646 & 1998ef & \nodata & \nodata & \nodata & \nodata & \nodata & 0.02 & 0.00 & 0.22 & \nodata & \nodata \\
NGC 0632 & 1998es & 3.14 & 1.53 & 0.19 & 0.71 & 0.30 & 0.76 & 6.20 & 2.53 & 1.29 & 1.01 \\
CGCG 180 022 & 1999X & \nodata & \nodata & \nodata & \nodata & \nodata & 0.01 & 0.03 & 0.03 & \nodata & \nodata \\
NGC 2595 & 1999aa & 2.05 & 0.14 & 0.07 & 0.10 & 0.00 & 0.37 & 2.73 & 1.48 & 0.73 & 0.67 \\
NGC 6063 & 1999ac & 1.30 & 0.42 & 0.00 & 0.14 & 0.07 & 0.20 & 1.73 & 0.80 & 0.41 & 0.64 \\
NGC 2841 & 1999by & \nodata & \nodata & \nodata & \nodata & \nodata & 0.09 & 0.56 & 1.25 & \nodata & \nodata \\
NGC 6038 & 1999cc & \nodata & \nodata & \nodata & \nodata & \nodata & 0.23 & 1.44 & 0.70 & \nodata & \nodata \\
NGC 6411 & 1999da & \nodata & \nodata & \nodata & \nodata & \nodata & 0.01 & 0.00 & 0.03 & \nodata & \nodata \\
NGC 6951 & 2000E & \nodata & \nodata & \nodata & \nodata & \nodata & 0.91 & 2.63 & 2.00 & \nodata & \nodata \\
\enddata
\end{deluxetable} 

\clearpage
\begin{deluxetable}{lccccccc} 
\tablewidth{0pt}
\tablenum{4}
\tabletypesize{\scriptsize}
\tablecaption{Galaxy Characterization\label{tbl-4}}
\tablecolumns{14}
\tablehead{\colhead{Galaxy} & \colhead{SN} & \colhead{M$_{B}$} & \colhead{$\delta$M$_{B}$}  &
\colhead{Morph. Type} & \colhead{H$\alpha$ Lum.} & \colhead{Log(O/H) + 12} & \colhead{Scalo \textit{b}} \cr \colhead{} &
\colhead{} & \colhead{(mag)} & \colhead{(mag)} & \colhead{} & \colhead{(ergs s$^{-1}$) } }
\startdata
NGC 4536 & 1981B & $-$19.86\tablenotemark{1} & 0.11 & Sbc & 6.87E+39 & 8.66 & 1.11 \\
NGC 3627 & 1989B & $-$20.5\tablenotemark{1} & 0.24 & Sb & 4.08E+39 & 9.13 & 0.57 \\
NGC 4639 & 1990N & $-$19.7\tablenotemark{1} & 0.23 & Sbc & 1.49E+39 & 8.76 & 0.37 \\
CGCG 111 016 & 1990O & \nodata & \nodata & Sba & 1.07E+40 & \nodata & 0.15 \\
NGC 4527 & 1991T & $-$20.5 & 0.27 & Sbc & 3.87E+39 & 8.83 & 0.35 \\
IC 4232 & 1991U & $-$21.02 & 0.32 & Sbc & 6.42E+40 & 9.34 & 0.82 \\
NGC 4374 & 1991bg & $-$21.33 & 0.07 & E1 & 9.42E+36 & \nodata & 0.01 \\
Anon & 1992J & $-$21.28 & 0.14 & S0$+$ & 1.79E+39 & \nodata & 0.01 \\
IC 3690 & 1992P & $-$19.91 & 0.27 & Sb & 1.20E+40 & \nodata & 0.32 \\
Anon & 1992ag & \nodata & \nodata & S\tablenotemark{2} & 6.97E+40 & 8.87 & 2.05 \\
Anon & 1992bp & $-$20.7 & 0.12 & E2/S0 & 2.89E+40 & \nodata & 0.35 \\
CGCG 307 023 & 1993ac & \nodata & \nodata & E & \nodata & \nodata & 0.01 \\
NGC 4526 & 1994D & $-$20.29\tablenotemark{3}  & 0.10 & S0 & \nodata & \nodata & 0.01 \\
NGC 4493 & 1994M & \nodata & \nodata & E+ pec & 2.01E+39 & \nodata & 0.01 \\
CGCG 224 104 & 1994Q & \nodata & \nodata & S0 & 1.99E+40 & 8.76 & 1.11 \\
NGC 4495 & 1994S & $-$19.94 & 0.30 & Sab & 2.90E+40 & 8.85 & 1.15 \\
NGC 3370 & 1994ae & $-$18.96 & 0.28 & Sc & 1.94E+39 & 8.79 & 0.95 \\
NGC 2962 & 1995D & $-$19.83 & 0.28 & Sa & 2.94E+37 & \nodata & 0.01 \\
NGC 2441 & 1995E & $-$20.77 & 0.28 & Sb & 5.34E+39 & 8.89 & 0.50 \\
IC 1844 & 1995ak & \nodata & \nodata & Sbc & 3.79E+40 & 8.79 & 1.19 \\
NGC 3021 & 1995al & $-$18.74 & 0.30 & Sbc & 7.36E+39 & 8.94 & 1.20 \\
UGC 03151 & 1995bd & \nodata & \nodata & S0 & 1.27E+40 & \nodata & 0.20 \\
Anon & 1996C & \nodata & \nodata & Sa & 1.16E+40 & 8.78 & 0.76 \\
NGC 2935 & 1996Z & $-$20.6 & 0.29 & Sb & 1.86E+39 & \nodata & 0.17 \\
Anon & 1996ab & \nodata & \nodata & \nodata & 8.24E+40 & \nodata & 0.14 \\
NGC 5005 & 1996ai & $-$19.72 & 0.29 & Sbc & 1.29E+39 & \nodata & 0.16 \\
Anon & 1996bl & \nodata & \nodata & SBc & 1.54E+40 & 8.64 & 0.43 \\
NGC 0673 & 1996bo & $-$21.27 & 0.31 & Sc & 3.55E+40 & 8.88 & 1.50 \\
UGC 03432 & 1996bv & $-$18.91 & 0.31 & Scd & 1.26E+40 & 8.46 & 0.46 \\
NGC 2258 & 1997E & $-$20.31 & 0.29 & S0 & \nodata & \nodata & 0.01 \\
NGC 4680 & 1997bp & \nodata & \nodata & Pec & 1.21E+40 & 8.96 & 1.11 \\
NGC 3147 & 1997bq & $-$21.59 & 0.28 & Sbc & 6.91E+39 & \nodata & 0.26 \\
Anon & 1997br & $-$17.68 & 0.32 & Sbd & 3.56E+39 & 8.80 & 1.53 \\
NGC 5490 & 1997cn & $-$21.33 & 0.30 & E & 9.10E+37 & \nodata & 0.01 \\
NGC 0105 & 1997cw & \nodata & \nodata & Sab & 2.07E+40 & 8.79 & 1.23 \\
UGC 03845 & 1997do & $-$19.55 & 0.30 & SBbc & 8.51E+39 & 8.62 & 0.92 \\
NGC 5440 & 1998D & $-$20.35 & 0.31 & Sa & 1.36E+38 & \nodata & 0.04 \\
NGC 6627 & 1998V & $-$20.5 & 0.39 & Sb & 3.46E+40 & 9.01 & 0.74 \\
NGC 4704 & 1998ab & $-$20.64 & 0.31 & Sbc pec & 2.07E+40 & 9.09 & 0.54 \\
NGC 3982 & 1998aq & $-$19.15 & 0.29 & Sb & 5.90E+39 & 8.92 & 1.41 \\
NGC 6495 & 1998bp & $-$20.02 & 0.30 & E & 1.75E+37 & \nodata & 0.01 \\
NGC 3368 & 1998bu & $-$20.21\tablenotemark{1} & 0.22 & Sab & 5.28E+38 & \nodata & 0.06 \\
NGC 0252 & 1998de & $-$20.84 & 0.28 & Sab & 1.05E+39 & \nodata & 0.04 \\
CGCG 302-013  & 1998di & \nodata & \nodata & \nodata & 2.92E+38 & \nodata & 0.02 \\
UGC 00139 & 1998dk & $-$19.43 & 0.35 & Sc & 1.10E+40 & 8.85 & 0.92 \\
UGCA 017 & 1998dm & \nodata & \nodata & SBc & 7.08E+39 & 8.49 & 1.06 \\
UGC 11149 & 1998dx & \nodata & \nodata & E2 & 9.66E+38 & \nodata & 0.01 \\
UGC 03576 & 1998ec & $-$20.22 & 0.31 & SBb & 6.18E+39 & \nodata & 0.18 \\
UGC 00646 & 1998ef & $-$19.48 & 0.30 & Sb & 0.00E+00 & \nodata & 0.02 \\
NGC 0632 & 1998es & $-$19.94 & 0.28 & S0\tablenotemark{4} & 1.20E+40 & 8.96 & 1.66 \\
CGCG 180-022 & 1999X & \nodata & \nodata & \nodata & 4.79E+38 & \nodata & 0.03 \\
NGC 2595 & 1999aa & $-$20.98 & 0.38 & Sc & 1.21E+40 & \nodata & 0.52 \\
NGC 6063 & 1999ac & $-$19.22 & 0.30 & Scd & 3.51E+39 & 8.90 & 0.44 \\
NGC 2841 & 1999by & $-$19.66 & 0.28 & Sb & 1.76E+38 & \nodata & 0.04 \\
NGC 6038 & 1999cc & $-$21.31 & 0.30 & Sc & 3.12E+40 & \nodata & 0.37 \\
NGC 6411 & 1999da & $-$20.67 & 0.28 & E & 0.00E+00 & \nodata & 0.01 \\
NGC 6951 & 2000E & $-$19.57 & 0.29 & Sbc & 1.24E+39 & \nodata & 0.15 \\
\enddata
\tablenotetext{1}{Gibson et al.(2000)}
\tablenotetext{2}{Interacting galaxy}
\tablenotetext{3}{Hamuy et al.(1996b)}
\tablenotetext{4}{Star burst core}
\end{deluxetable}

\clearpage
\begin{deluxetable}{lcccc}
\tablewidth{0pt}
\tablenum{5}
\tabletypesize{\scriptsize}
\tablecaption{Type Ia Supernova Characterization\label{tbl-5}}
\tablehead{\colhead{Galaxy} & \colhead{SN} & 
\colhead{$\Delta$m$_{15}$(B)} & 
\colhead{$\sigma$} & \colhead{Hubble Residual} \cr \colhead{} & \colhead{} & \colhead{(mag)} 
& \colhead{(mag)} & \colhead{(mag)}}
\startdata
NGC 4536 & 1981B & 1.10 & 0.07 & \nodata \\
NGC 3627 & 1989B & 1.31 & 0.07 & \nodata \\
NGC 4639 & 1990N & 1.07 & 0.05 & \nodata \\
CGCG 111 016 & 1990O & 0.96 & 0.10 & 0.11 \\
NGC 4527 & 1991T & 0.94 & 0.05 & \nodata \\
IC 4232 & 1991U & 1.06 & 0.10 & $-$0.31 \\
NGC 4374 & 1991bg & 1.93 & 0.10 & \nodata \\
Anon & 1992J & 1.56 & 0.10 & $-$0.23 \\
IC 3690 & 1992P & 0.87 & 0.10 & 0.20 \\
Anon & 1992ag & 1.19 & 0.10 & $-$0.13 \\
Anon & 1992bp & 1.32 & 0.10 & $-$0.12 \\
CGCG 307 023 & 1993ac & 1.19 & 0.10 & 0.06 \\
NGC 4526 & 1994D & 1.32 & 0.05 & \nodata \\
NGC 4493 & 1994M & 1.44 & 0.10 & $-$0.19 \\
CGCG 224-104 & 1994Q & 1.03 & 0.10 & $-$0.12 \\
NGC 4495 & 1994S & 1.10 & 0.10 & 0.03 \\
NGC 3370 & 1994ae & 0.86 & 0.05 & \nodata \\
NGC 2962 & 1995D & 0.99 & 0.05 & \nodata \\
NGC 2441 & 1995E & 1.06 & 0.05 & \nodata \\
IC 1844 & 1995ak & 1.26 & 0.10 & $-$0.34 \\
NGC 3021 & 1995al & 0.83 & 0.05 & \nodata \\
UGC 03151 & 1995bd & 0.84 & 0.05 & $-$0.13 \\
Anon & 1996C & 0.97 & 0.10 & 0.32 \\
NGC 2935 & 1996Z & 1.22 & 0.10 & \nodata \\
Anon & 1996ab & 0.87 & \nodata & 0.23 \\
NGC 5005 & 1996ai & 0.99 & 0.10 & \nodata \\
Anon & 1996bl & 1.17 & 0.10 & 0.12 \\
NGC 0673 & 1996bo & 1.25 & 0.05 & \nodata \\
UGC 03432 & 1996bv & 0.93 & 0.10 & $-$0.15 \\
NGC 2258 & 1997E & 1.39 & 0.06 & 0.14 \\
NGC 4680 & 1997bp & 1.00 & 0.05 & $-$0.16 \\
NGC 3147 & 1997bq & 1.00 & 0.05 & $-$0.08 \\
Anon & 1997br & 1.02 & 0.06 & \nodata \\
NGC 5490 & 1997cn & 1.90 & 0.05 & 0.16 \\
NGC 0105 & 1997cw & 1.02 & 0.10 & $-$0.07 \\
UGC 03845 & 1997do & 0.99 & 0.10 & 0.24 \\
NGC 5440 & 1998D & \nodata & \nodata & 0.02 \\
NGC 6627 & 1998V & 1.06 & 0.05 & $-$0.03 \\
NGC 4704 & 1998ab & 0.88 & 0.17 & $-$0.12 \\
NGC 3982 & 1998aq & 1.14 & \nodata & \nodata \\
NGC 6495 & 1998bp & 1.83 & 0.06 & $-$0.05 \\
NGC 3368 & 1998bu & 1.01 & 0.05 & \nodata \\
NGC 0252 & 1998de & 1.93 & 0.05 & 0.02 \\
CGCG 302 013  & 1998di & \nodata & \nodata & \nodata \\
UGC 00139 & 1998dk & 1.05 & 0.10 & 0.06 \\
UGCA 017 & 1998dm & 1.07 & 0.06 & \nodata \\
UGC 11149 & 1998dx & 1.55 & 0.09 & $-$0.08 \\
UGC 03576 & 1998ec & 1.08 & 0.09 & 0.09 \\
UGC 00646 & 1998ef & 0.97 & 0.10 & \nodata \\
NGC 0632 & 1998es & 0.87 & 0.08 & $-$0.06 \\
CGCG 180 022 & 1999X & 1.11 & 0.08 & 0.04 \\
NGC 2595 & 1999aa & 0.85 & 0.08 & \nodata \\
NGC 6063 & 1999ac & 1.00 & 0.08 & 0.17 \\
NGC 2841 & 1999by & 1.87 & \nodata & \nodata \\
NGC 6038 & 1999cc & 1.46 & 0.05 & 0.05 \\
NGC 6411 & 1999da & 1.94 & \nodata & 0.10 \\
NGC 6951 & 2000E & 0.94 & 0.50 & \nodata \\
\enddata
\end{deluxetable}


\begin{thebibliography}{}
\bibitem[Andrievsky et al.(2004)]{and04}Andrievsky, S. M., Luck, R. E., Martin, P., \& L\'{e}pine, 
    J. R. D. 2004, \aap, 413, 159
\bibitem[Caldwell \& Oemler(1981)]{Cald81}Caldwell, C. N., Oemler, A. 1981, \aj, 86, 1424
\bibitem[Dominguez et al.(1999)]{Dom99}Dominguez, I., Chieffi, A., \& Straniero, O. 1999
    , \apj, 524, 226
\bibitem[Fabricant et al.(1998)]{Fab98}Fabricant, D., Cheimets, P., Caldwell, N., \& Geary, J. 
    1998, \pasp, 110, 79
\bibitem[Filippenko(1982)]{Fil82}Filippenko, A. V. 1982, \pasp, 94, 715
\bibitem[Freedman et al.(2001)]{Fre01}Freedman, W. L., et al. 2001, \apj, 533, 47F
\bibitem[Gal-Yam \& Maoz(2004)]{GalYam}Gal-Yam, A., Maoz, D. 2004, \mnras, 347, 942
\bibitem[Garnavich et al.(1998)]{G98}Garnavich, P. M., et al. 1998, \apj, 509, 74G
\bibitem[Garnavich et al.(2004)]{garn04}Garnavich, P. M. et al. 2004, \apj, 613, 1120
\bibitem[Geller et al.(1997)]{Gell97}Geller, M., et al. 1997, \aj, 114, 2205
\bibitem[Gibson et al.(2000)]{Gib00}Gibson, B. K., et al. 2000, \apj, 529, 723
\bibitem[Hamuy et al.(1996b)]{H96b}Hamuy, M., Phillips, M. M., Suntzeff, N. B.,
    Schommer, R. A., Maza, J., Aviles, R. 1996b, \aj, 112, 2391
\bibitem[Hamuy et al.(1996c)]{H96c} Hamuy, M., Phillips, M. M., Suntzeff, N. B.,
    Schommer, R. A., Maza, J., Aviles, R. 1996c,\aj, 112, 2398set out 
\bibitem[Hamuy et al.(2000)]{H00}Hamuy, M., et al. 2000, \apj, 120, 1479
\bibitem[Henry \& Worthey(1999)]{HW99}Henry, R. B. C., \& Worthey, G.  1999, \pasp, 111, 919
\bibitem[H\"{o}flich et al.(1998)]{HWT98}H\"{o}flich, P., Wheeler, J. C.,
    \& Thielemann, F. K. 1998, \apj, 495, 617 
\bibitem[H\"{o}flich et al.(2000)]{HNUW00}H\"{o}flich, P., Nomoto, K.,
    Umeda, H., \& Wheeler, J. C. 2000, \apj, 528, 590  
\bibitem[H\"{o}flich et al.(2002)]{Hof03}H\"{o}flich, P., Gerardy, C. L., Fesen, R. A., Sakai, S. 2002,
    \apj, 568, 791
\bibitem[Iben \& Tutukov(1984)]{IT84}Iben, I. Jr. \& Tutukov, A. V. 1984, \apjs, 54, 355
\bibitem[Jansen et al.(2000)]{jansen}Jansen, R. A.,
    Fabricant, D., Franx, M., \& Caldwell, N. 2000, \apjs, 126, 331
\bibitem[Jha et al.(1999)]{Jha99}Jha, S., et al. 1999, \apjs, 125, 73
\bibitem[Jha(2002)]{Jha02}Jha, S. 2002, PhD Thesis 
\bibitem[Kennicutt, Tamblyn, \& Congdon(1994)]{KTC94}Kennicutt, R. C., Tamblyn, P., \& Congdon, C. W.
    1994, \aj, 435, 22
\bibitem[Kewley \& Dopita(2002)]{KD02}Kewley, L. J., Dopita, M. A. 2002, \apjs, 142, 35
\bibitem[Khokhlov(1991)]{Klv91}Khokhlov, A. M. 1991, \aap, 245, 114
\bibitem[Krisciunas et al.(2004)]{kevin}Krisciunas, K., Phillips, M. M., Suntzeff, N. B. 2004, \apj, 602, L81
\bibitem[Kroupa et al.(1993)]{ktg93}Kroupa, P., Tout, C. A., \& Gilmore, G. 1993, \mnras, 262, 545
\bibitem[Lanzetta et al.(2002)]{lanz02}Lanzetta, K. M., Yahata, N., Pascarelle, S., Chen, H., Fern$\acute{a}$ndez-Soto, A.,
   2002, \apj, 570, 492
\bibitem[Madau et al.(1998b)]{madau98}Madau, P., et al. 1998b, \mnras, 498, 106
\bibitem[Mannucci et al.(2005)]{mann05} Mannucci, F., et al.  2005, \aap, 433, 807
\bibitem[Massey et al.(1988)]{Mass88}Massey, P., Strobel, K., Barnes, J. V., 
    \& Anderson, E. 1988, \apj, 328, 315
\bibitem[McMillan \& Ciardullo(1996)]{MC96}McMillan, R. J., Ciadullo, R. 1996, \apj, 473, 707M
\bibitem[Miknaitis et al. (2005)]{gajus05} Miknaitis, G. et al. 2004, in press
\bibitem[Miller \& Mathews(1972)]{MM72}Miller, M. S., Mathews, W. G. 1972, \apj, 172, 593
\bibitem[Nomoto(1982)]{Nom82}Nomoto, K. 1982, \apj, 253, 798
\bibitem[Oemler \& Tinsley(1979)]{Oeml79}Oemler, A., Tinsley, B. M. 1979, \aj, 84, 9850
\bibitem[Perlmutter et al.(1997)]{Perl97}Perlmutter, S., et al. 1997, \apj, 483, 565P
\bibitem[Perlmutter et al.(1999)]{Perl99}Perlmutter, S., et al. 1999, \apj, 517, 565
\bibitem[Phillips(1993)]{P93} Phillips, M. M. 1993,\apj, 413, L105
\bibitem[Phillips et al.(1999)]{Phil99}Phillips, M. M., Lira, P., Suntzeff, N. B., Schommer, R. A.,
    Hamuy, M., Maza, J. 1999,\aj, 118, 1766
\bibitem[Qian \& Wasserburg(2001)]{QW01}Qian, Y. Z., Wasserburg, G. J., 2001, \apj, 549, 337
\bibitem[Reiss, Press, \& Kirshner(1996)]{RPK96}Reiss, A. G., Press, W. H., Kirshner, R. P.
    1996, \apj, 473, 88
\bibitem[Riess et al.(1998)]{R98}Riess, A. G., et al. 1998, \aj, 116, 1009R
\bibitem[Riess et al.(1999)]{Rie99}Riess, A. G., et al. 1999, \aj, 117, 707
\bibitem[Scalo(1986)]{Sc86}Scalo, J. M. 1986, \fcp, 11, 1
\bibitem[Schechter(1976)]{Schec76}Schechter, P. 1976, \apj, 203, 297
\bibitem[Schmidt et al.(1998)]{Schmidt98}Schmidt, B. P., et al. 1998, \apj, 507, 46S
\bibitem[Strolger et al.(2003)]{Strol03} Strolger, L-G. et al. in press
\bibitem[Timmes, Brown, \& Truran(2003)]{TBT03}Timmes, F. X., Brown, E. F., Truran, J. W.
    2003, \apj, 590, L83
\bibitem[Umeda et al.(1999)]{Um99} Umeda, H., Nomoto, K., Kobayashi, C., Hachisu, I., 
    \& Kato, M. 1999, \apj, 522, L43
\bibitem[Wang \& Garnavich (2001)]{wang01} Wang, Y., \& Garnavich, P. M. 2001, \apj, 552, 445
\bibitem[Webbink(1984)]{Webb84}Webbink, R. F. 1984, \apj, 227, 355
\bibitem[Whelan \& Iben(1973)]{WI73}Whelan, J. \& Iben, I. Jr. 1973, \apj, 186, 1007
\bibitem[Woosely \& Weaver(1994a)]{WW94}Woosley, S. E., \& Weaver, T. A. 1994, 
    in Supernovae (Amsterdam: Elsevier), 423
\bibitem[Worthey(1994)]{Worthey94}Worthey, G. 1994, \apj, 95, 107
\bibitem[Yamaoka et al.(1992)]{Yam92}Yamaoka, H. Nomoto, K., Shigeyama, T., \& Thielemann,
    F. 1992, \apj, 393, L55
\bibitem[Harvard Supernova List()]{SupLst} http://cfa-www.harvard.edu/iau/lists/Supernovae.html
\bibitem[CBAT()]{cbat}http://cfa-www.harvard.edu/iau/cbat.html
\end{thebibliography}
\end{document}